\documentclass[review]{elsarticle}
\usepackage{tabularx}
\usepackage{amsmath}
\usepackage{lineno,hyperref}
\usepackage{adjustbox,lipsum}
\modulolinenumbers[5]
\journal{Journal of Advances in High Energy Physics}

\begin{document}
\begin{frontmatter}
\title{Compelling evidence of oscillatory behaviour of hadronic multiplicities in the shifted Gompertz distribution}

\author{R.~Aggarwal}
\ead{ritu.~aggarwal1@gmail.com}
\address{Savitribai Phulle Pune University, Pune, India-140007}
\author{M.~Kaur\fnref{myfootnote}}
\ead{manjit@pu.ac.in}
\address{Physics Department, Panjab University, Chandigarh, India-160014 }
 
\begin{abstract}
Study of charged particle multiplicity distribution in high energy interactions of particles helps in revealing the dynamics of particle production and the underlying statistical patterns, which these distributions follow.~Several distributions derived from statistics have been employed to understand its behaviour.~In one of our earlier papers, we introduced the shifted Gompertz distribution to investigate this variable and showed that the multiplicity distributions in a variety of processes at different energies can be very well described by this distribution.~The fact that the shifted Gompertz distribution, which has been extensively used in diffusion theory, social networks and forecasting has been used for the first time in high energy physics collisions, remains interesting.~In this paper we investigate the phenomenon of oscillatory behaviour of the counting statistics observed in the high energy experimental data, resulting from different types of recurrence relations defining the probability distributions.~We search for such oscillations in the multiplicity distributions well described by the shifted Gompertz distribution and look for retrieval of additional valuable information from these distributions. 

\end{abstract}


\end{frontmatter}

\section{Introduction}
The simplest observable in high energy interactions, is a count of charged particles produced in a collision and its mean value.~Its distribution measured in full or partial phase space forms both a tool for studying models and probe for particle dynamics.~A large number of statistical probability distribution functions (PDF) have been used to understand its behaviour.~These include Koba, Nielsen and Olesen (KNO) scaling \cite{KNO}, Poisson distribution \cite{POI}, binomial and negative binomial \cite{NBD} distributions, lognormal distribution \cite{LOG}, Tsallis distribution \cite{TS1, TS2}, Weibull distribution \cite{WEI}, modified forms of these and several other distributions.~NBD has been one of the most extensively used.~It was very successful until the results from UA5 collaboration \cite{UA51, UA52} published.~A shoulder structure was observed in the multiplicity distribution in $p\overline{p}$ collisions, showing its violations.~It is also well established by various experimental results that NBD fails with increasing deviations with the growing number of charged particles produced.~In order to improve the agreement with data, 2-component or 3-component NBD fits \cite{ZBO, ZBO2} were also used.

In one of our recent papers, we introduced the shifted Gompertz distribution \cite{shGomp}, henceforth named as SGD, to investigate the multiplicities in various leptonic and hadronic collisions over a large range of collision energies.~The distribution was first introduced by Bemmaor \cite{BEMA} as a model of adoption of innovations.~The two non-negative fit parameters define the scale and shape of the distribution.~This distribution has been widely studied in various contexts \cite{Jon, Jod, Jim}.~In our earlier work \cite{shGomp} we proposed to use the SGD for studying the charged particle multiplicities in high energy particle collisions.~And showed from a detailed study for collisions in full phase space and also in limited phase space that this distribution explained the experimental data very well in high energy particle collisions using leptons and hadrons as probes.~Subsequently we also used it to calculate the higher moments of a multiplicity distribution which also serve as a powerful tool to unfold the characteristics and correlations of particles \cite{AM}.~We also used 2-component shifted Gompertz distribution, named as modified shifted Gompertz (MSGD) to successfully improve the agreement between data and fit.~The details are given in our paper \cite{shGomp}.

Wilk and  W{\l}odarczyk, in one of their recent publications \cite{WILK}, pointed out that the 2-component or multi-component fits improve the agreement only at large $N$ (number of charged particles) but not at small $N$.~They showed that the ratio $data/fit$ deviates significantly from unity for small $N$.~In a pursuit of retrieving additional information from measured probability of producing $N$ particles $P(N)$, they have proposed the multiplicity distribution (MD) by a recurrence relation between the adjacent distributions $P(N)$ and $P(N+1)$.~This corresponds to the assumption of a connection existing only between the production of $N$ and $N+1$ particles:
\begin{equation}
(N+1)P(N+1) = g(N)P(N).
\end{equation}
The multiplicity distribution is then determined by the function form of $g(N)$, the simplest being a linear relation:
\begin{equation}
g(N) = \mu + \nu N.
\end{equation}
where $\mu$ and $\nu$ are the parameters of the linear dependence.
The more general form of recurrence relation introduced in reference \cite{WILK} which connects the multiplicity $N + 1$ with all smaller multiplicities has the form; 
\begin{equation}
(N + 1)P(N + 1) = <N>\sum_{j=0}^{N} C_{j}P(N-j).
\end{equation}
All multiplicities are then connected by means of some coefficients $C_j$, which redefine the corresponding $P(N)$ in the way such that the coefficients $C_j$ connect the probability of particle $N+1$ with probabilities of all the $N-j$ previously produced particles.~These coefficients can then be directly calculated from the experimentally measured $P(N)$ by exploiting the relationship.~It is shown that the $C_j$ shows a very distinct oscillatory behaviour which gradually diminishes with increasing $N$ and nearly vanishes.~The details are given in Section~3.

In the present work we use shifted Gompertz distribution and its modified forms using the data at high energies from $p\overline{p}$ and $pp$ interactions to understand the existence of such oscillatory behaviour and to check if we obtain the results consistence with the ones from \cite{WILK}.

In Section~2, we provide the essential formulae for the Probability Distribution Function of the shifted Gompertz distribution and modified 2-component shifted Gompertz distributions, in brief.~A very brief description of the how the oscillations have been estimated in the multiplicity distributions by Wilk et al \cite{WILK} is included for the sake of completeness.

Section~3 presents the analysis of experimental data, the fitted shifted Gompertz distributions, the fitted modified shifted Gompertz and the distributions giving out the oscillatory behaviour.~Discussion and conclusion are presented in Section~4. 

\section{Shifted Gompertz distribution (SGD)}

Let X be any non-negative random variable having the shifted Gompertz distribution with parameters $b$ and $\zeta$, where b $>$ 0 is a scale parameter and $\zeta >$ 0 is a shape parameter.~Value of the scale parameter, determines the statistical dispersion of the probability distribution.~Larger the value of the scale parameter, more is the distribution spread out and smaller the value, the distribution being more concentrated.~The shifted Gompertz density function can take on different shapes depending on the values of the shape parameter $\zeta$.~It is a kind of numerical parameter which affects the shape of a distribution rather than simply shifting it or stretching or shrinking it.~The multiplicity distribution is measured as the probability distribution of a number of particles being produced in a collision at a particular energy of collision and follows certain phenomenological and statistical models.~The probability distribution function of X is given by
\begin{equation}
P_X(x;b,\zeta) = b e^{-(bx + \zeta e^{-bx})}\big(1+\zeta(1-e^{-bx})\big), \>\>\>\>where\>\>\>x>0
\end{equation}

The Mean value (E[X]) of Shifted Gompertz distribution is given by 
\begin{equation}
E[X] = \frac{1}{b}( \gamma + log[\zeta] + \frac{1 - e^{-\zeta}}{\zeta} + \Gamma[0,\zeta] )
\end{equation}
where $\gamma$ $\approx$ 0.5772156 stands for the Euler constant (also referred to as Euler-Mascheroni constant).
It is well established that at high energies the most widely adopted, Negative Binomial distribution \cite{NBD} fails and deviates significantly for high multiplicity tail, from the experimental data.~To extend the applicability of NBD, another approach, was introduced by A. Giovannini et al \cite{NBD}.~In this case a weighted superposition of two independent NBDs, one corresponding to the soft events (events without mini-jets) and another to the semi-hard events (events with mini-jets), is obtained.~These distributions combine merely two classes of events and not two different particle-production mechanisms.~We used the same method to obtain the superposed distribution and call it 2-component shifted Gompertz distribution (2-component SGD) as given by equation~(6).~The multiplicity distribution of each component being independent SGD.~The concept of superposition originates from purely phenomenological considerations.~The two fragments of the distribution suggest the presence of the substructure.~Each component-distribution has two fit parameters, namely scale and shape parameters.~The best fit overall distribution to the experimental data, with optimised parameters, also gives an estimate of fraction, $\alpha$, of the soft collisions, at a given c.m.s energy.~The dynamics of particle production is understood in terms of weighted superposition of soft and semi-hard contributions.~Though these superimposed physical substructures are different, the weighted superposition mechanism is the same.~The physical substructures are described by the same SGD multiplicity distributions and corresponding correlation functions, which are QCD inspired genuine self-similar fractal processes \cite{NBD}. Same as NBD, SGD allows to describe the multiplicity distribution on purely phenomenological grounds.~This may help in differentiating between different phenomenological models.~Details are included in our earlier publication \cite{shGomp}. 

\small
\begin{equation}
P_N(\alpha: b_1, \zeta_1; b_2, \zeta_2)=\alpha P_{soft}^{shGomp}(N)+(1-\alpha)P_{semi-hard}^{shGomp}(N)
\end{equation} 
\normalsize
where $\alpha$ is the fraction of soft events, ($b_1$, $\zeta_1$) and ($b_2$, $\zeta_2$) are respectively the scale and shape parameters of the two distributions.
\normalsize
\subsection{Modified forms of Shifted Gompertz distribution }

In this paper we adopt a different approach and investigate what kind of changes in the structure of the multiplicity distribution described by the SGD, are necessary in order to describe the same data by a single SGD, with accordingly modified parameters $b$ and $\zeta$.~To describe data using only a single SGD, we allow the parameter $b$ to depend on the multiplicity $N$, as suggested by Wilk et al \cite{WILK}.~To obtain an exact fit of the distribution to the experimental data, a non-monotonic dependence of $b$ on $N$ is introduced.~This way, the scale parameter $b$ remains the same in nature, but varies in accordance with the number of particles produced.~Such a change means that we preserve the overall form of the SGD;
\begin{equation}
b = b(N) = c \exp(a_1|N - d|)
\end{equation}
where $a_1$, $d$ and $c$ are parameters.~This leads to the modification of SGD (equation(4)) which describes the data very well.~We call this as the modified-SGD1 (MSGD1).~When another non-linear term with a coefficient $a_2$ is added \cite{WILK} to bring improved agreement with the data;
\begin{equation}
b = c\exp[(a_1|N - d|) + (a_2|N - d|)^4]
\end{equation}
~we call this second modification as MSGD2.~Further, we investigate the possibility of retrieving some additional information from the measured $P(N)$.

\section{Analysis and Results} 
The equation~(3) can be reversed and a recurrence formula can be obtained for the coefficients $C_j$ for an experimentally measured multiplicity distribution $P(N)$, as below;
\begin{equation}
\langle N \rangle C_j=(j+1)\left[\frac{P(j+1)}{P(0)}\right] - \langle N\rangle \sum_{i=0}^{j-1} C_{i}\left[ \frac{P(j-i)}{P(0)}\right]. 
\end{equation}
The errors on the coefficients $C_{j}$ are calculated from the variance;
\begin{equation}
\begin{aligned}
Var[\langle N \rangle C_j]&=\left[\frac{(j+1)}{P(0)}\right]^2 Var\left[P(j+1)\right]\\
 & + \sum_{i=0}^{j-1} (\langle N\rangle C_{i})^2 Var\left[P(j-i)\right]\\ 
  & + \sum_{i=0}^{j-1}\left[\frac{P(j-i)}{P(0)}\right]^2 Var[\langle N \rangle C_i].
\end{aligned}
\end{equation}
Since the coefficients $C_{j}$ are correlated, the last term of equation~(10) introduces dependence
of the error in $C_{j}$ on the errors of all coefficients with $i<j$.~This leads to a cumulative effect with a large increase of errors with increasing rank $j$.~However, despite such large errors, the values of $\langle N \rangle C_j$ lie practically on the curve and the points do not oscillate in the limits of errors.~Hence, the errors can be estimated reasonably well, by neglecting this cumulative effect. 

In the present work, calculations are performed using the data from different experiments and following two collision types;\\
i) $pp$ collisions at LHC energies $\sqrt{s}$ = 900, 2360 and 7000~GeV \cite{CMS} are analysed in five rapidity windows, $|\eta|$ $<$0.5 up to $|\eta|$ $<$ 2.4,\\
ii) $p\overline{p}$ collisions at energies $\sqrt{s}$ = 200, 540 and 900~GeV \cite{UA51,UA52} are analysed in full phase space as well as in rapidity windows, $|\eta|$ $<$0.5 up to $|\eta|$ $<$ 5.0.

The charged hadron multiplicity experimental distributions are fitted with the SGD~(equation~4), the 2-component SGD(equation~6), MSGD1 (equation~7) and MSGD2 (equation~8) for all rapidity windows at all energies.~It is observed that data do not show good agreement with fits for the lower and for very high values of $N$ with SGD.~However, the agreement becomes very good in both the limits when 2-component fits are performed.~A further improvement is shown with MSGD1 and MSGD2 fits in almost every case.~To  avoid a multitude of similar figures, we only show the probability distributions in figure~1, at $\sqrt{s}$=7000, 2360 and 900~GeV for $pp$ collisions in one rapidity window, $|\eta|< $ 2.4.~The fitted curves correspond to the distributions, the SGD, the 2-component SGD, MSGD1 and MSGD2.~For comparison between different fits, table~1 gives the $\chi^2/ndf$ for all fits at different energies and rapidities.~In case of 2-component SGD, the $\alpha$ values are taken from the reference \cite{shGomp}.

\begin{figure}[ht]
\includegraphics[width=4.8 in,height =2.43 in]{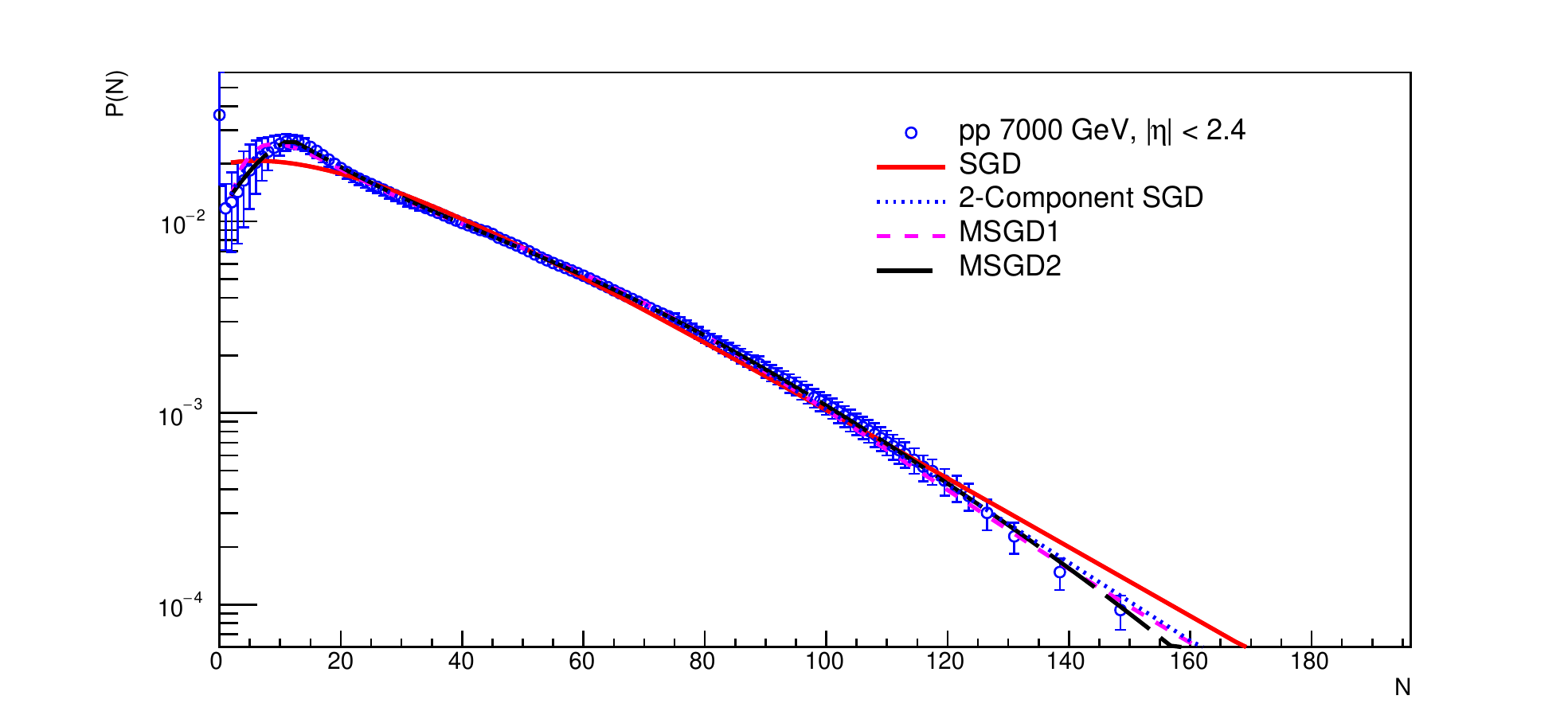}
\includegraphics[width=4.8 in,height =2.43 in]{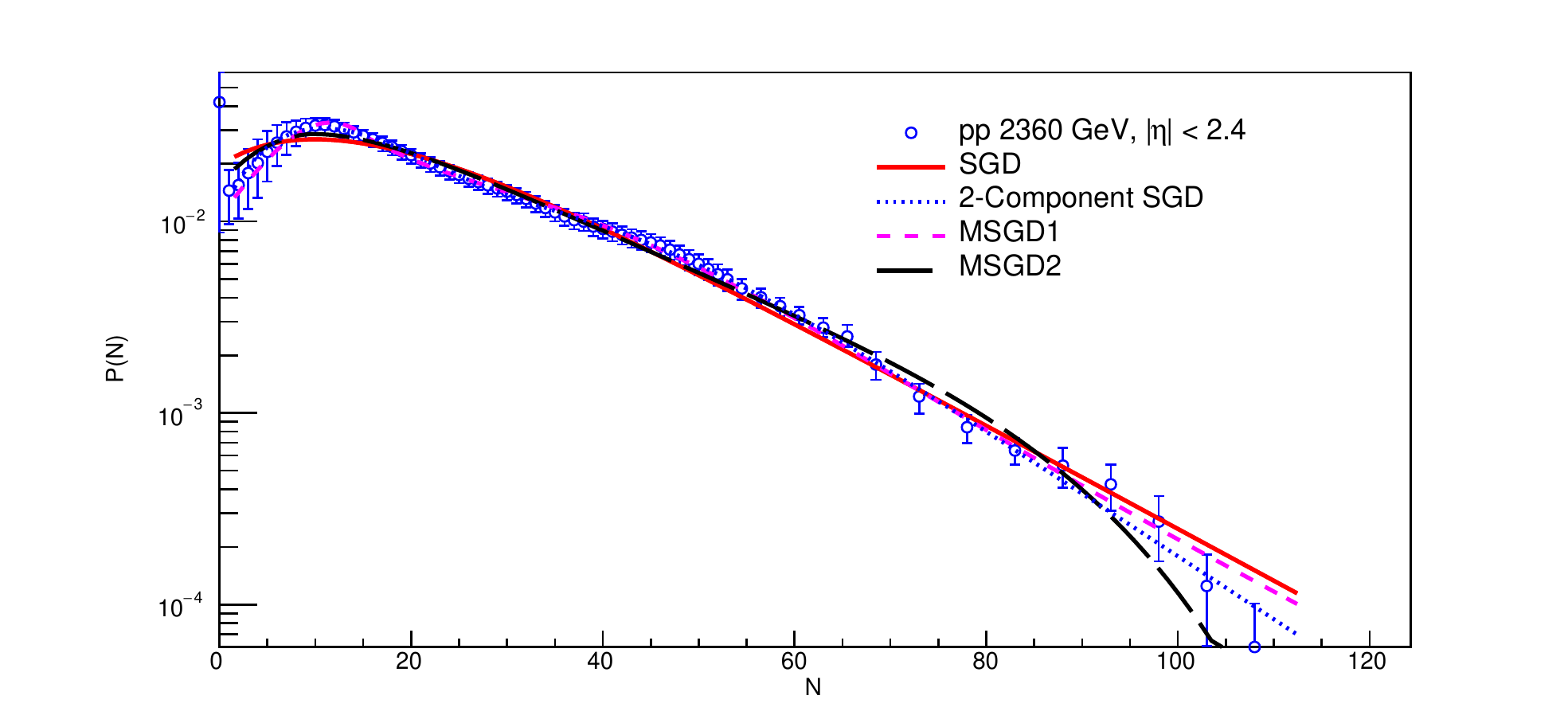}
\includegraphics[width=4.8 in,height =2.43 in]{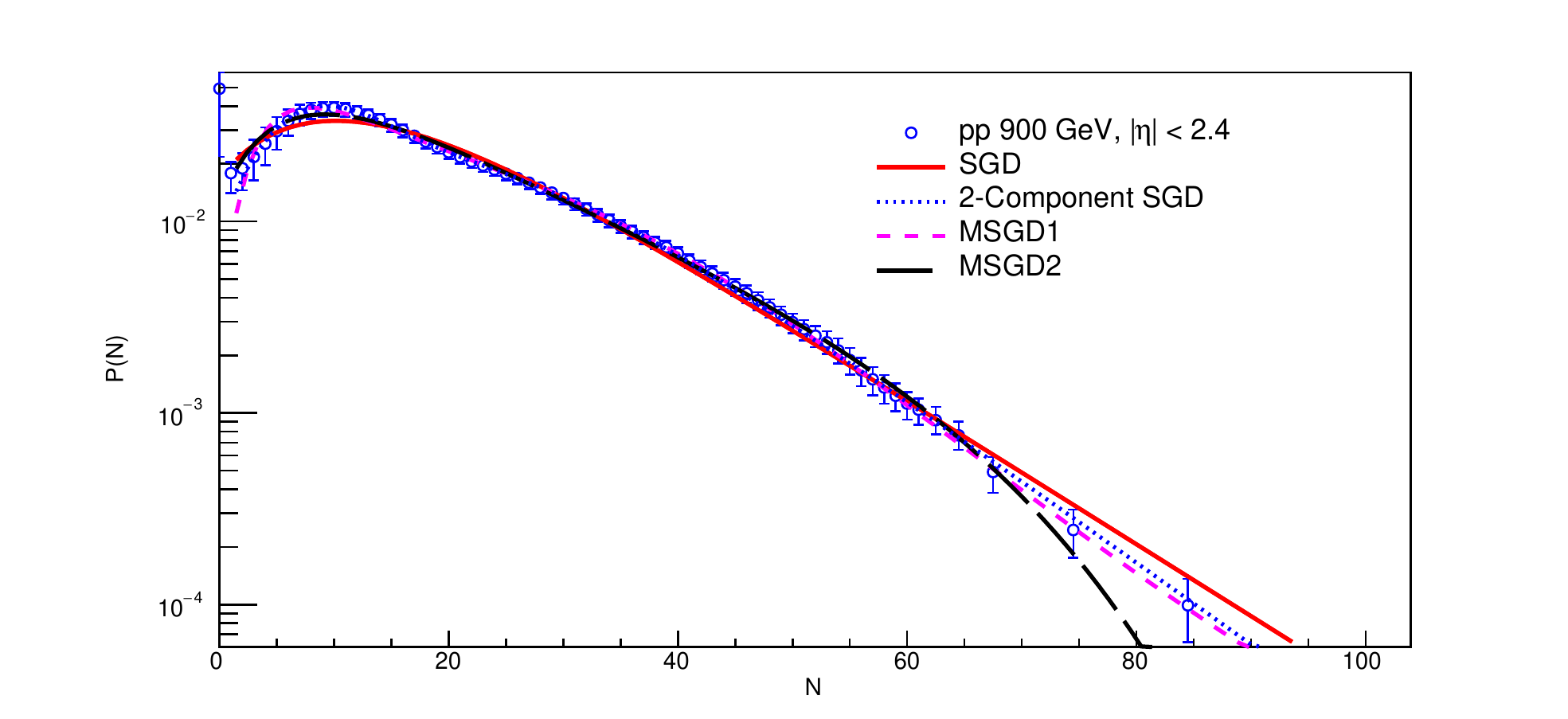}
\tiny
\caption{SGD, 2-component SGD, MSGD1 and MSGD2 distributions for data on $pp$ collisions at different $\sqrt{s}$ obtained by the CMS experiment for $|\eta| <2.4$}
\end{figure}

\begin{table*}[t]
\small
\begin{tabular}{|c|c|c|c|c|c|c|c|}
\hline
Energy   & $Rapidity   $ & $  SGD  $ &  2-component &  $   MSGD1   $ &  $  MSGD2  $ \\ 
$(GeV)$  &  interval           &           &  $SGD$      &                &              \\ \hline           
  & $|\eta|<$ & $\chi^{2}/ndf$ &  $\chi^{2}/ndf$ &  $\chi^{2}/ndf$ &  $\chi^{2}/ndf$ \\ \hline
          
 900 & 0.5 &      3.57 / 19 &      0.79 / 16 &      0.97 / 17 &       0.57 / 16 \\ \hline 
 900 & 1.0 &     17.50 / 32 &     11.16 / 29 &      3.41 / 30 &       6.32 / 29 \\ \hline 
 900 & 1.5 &     66.98 / 48 &     12.59 / 45 &     13.74 / 46 &      11.02 / 45 \\ \hline 
 900 & 2.0 &     55.41 / 58 &      8.17 / 55 &     19.38 / 56 &      17.27 / 55 \\ \hline 
 900 & 2.4 &     72.26 / 64 &     12.63 / 61 &     21.79 / 62 &      22.32 / 61 \\ \hline 
 2360 & 0.5 &      8.13 / 19 &      2.75 / 16 &      5.41 / 17 &       4.24 / 16 \\ \hline 
 2360 & 1.0 &     24.30 / 34 &     22.99 / 31 &     15.32 / 32 &       7.55 / 31 \\ \hline 
 2360 & 1.5 &     28.08 / 45 &      3.74 / 42 &      7.51 / 43 &       6.02 / 42 \\ \hline 
 2360 & 2.0 &     39.83 / 55 &     22.71 / 52 &      9.77 / 53 &       9.76 / 52 \\ \hline 
 2360 & 2.4 &     59.55 / 66 &      7.85 / 63 &     17.34 / 64 &      33.03 / 63 \\ \hline 
 7000 & 0.5 &    117.47 / 37 &     13.50 / 34 &      8.28 / 35 &       8.49 / 34 \\ \hline 
 7000 & 1.0 &    223.71 / 66 &     27.11 / 63 &     28.33 / 64 &      13.27 / 63 \\ \hline 
 7000 & 1.5 &    247.86 / 88 &     26.46 / 85 &     88.09 / 86 &       7.62 / 85 \\ \hline 
 7000 & 2.0 &    164.61 / 108 &     25.09 / 105 &     35.37 / 106 &      10.17 / 105 \\ \hline 
 7000 & 2.4 &    179.74 / 123 &     27.45 / 120 &     33.91 / 121 &       5.57 / 120 \\ \hline 

\end{tabular}
\caption{$\chi^{2}/ndf$ for charged multiplicity distribution fitted with Shifted Gompertz, 2-component Shifted Gompertz, MSGD1 and MSGD2 distributions for $pp$ collisions.}
\end{table*}

Figure~2 shows the similar distributions at $\sqrt{s}$=900, 540 and 200~GeV for $p\overline{p}$ collisions in one rapidity window $|\eta|<$ 3.0.~The fitted curves correspond to the distributions, the SGD, the 2-component SGD, MSGD1 and MSGD2.~For comparison between different fits, table~2 gives the $\chi^2/ndf$ for all fits at different energies and rapidities.~It may be observed in the cases of $pp$ collisions, MSGD2 fits the data well in comparison to other distributions, particularly at higher energies.~However for the case of $p\overline{p}$ collisions, in most of the cases 2-component SGD improves the fits and explains the data well.~A comparison between $pp$ and $p\overline{p}$ collsions at the same $\sqrt{s}$=900 GeV, the trend is nearly the same and MSGD2 fit the data best.

\begin{figure}[ht]
\includegraphics[width=4.8 in,height =2.43 in]{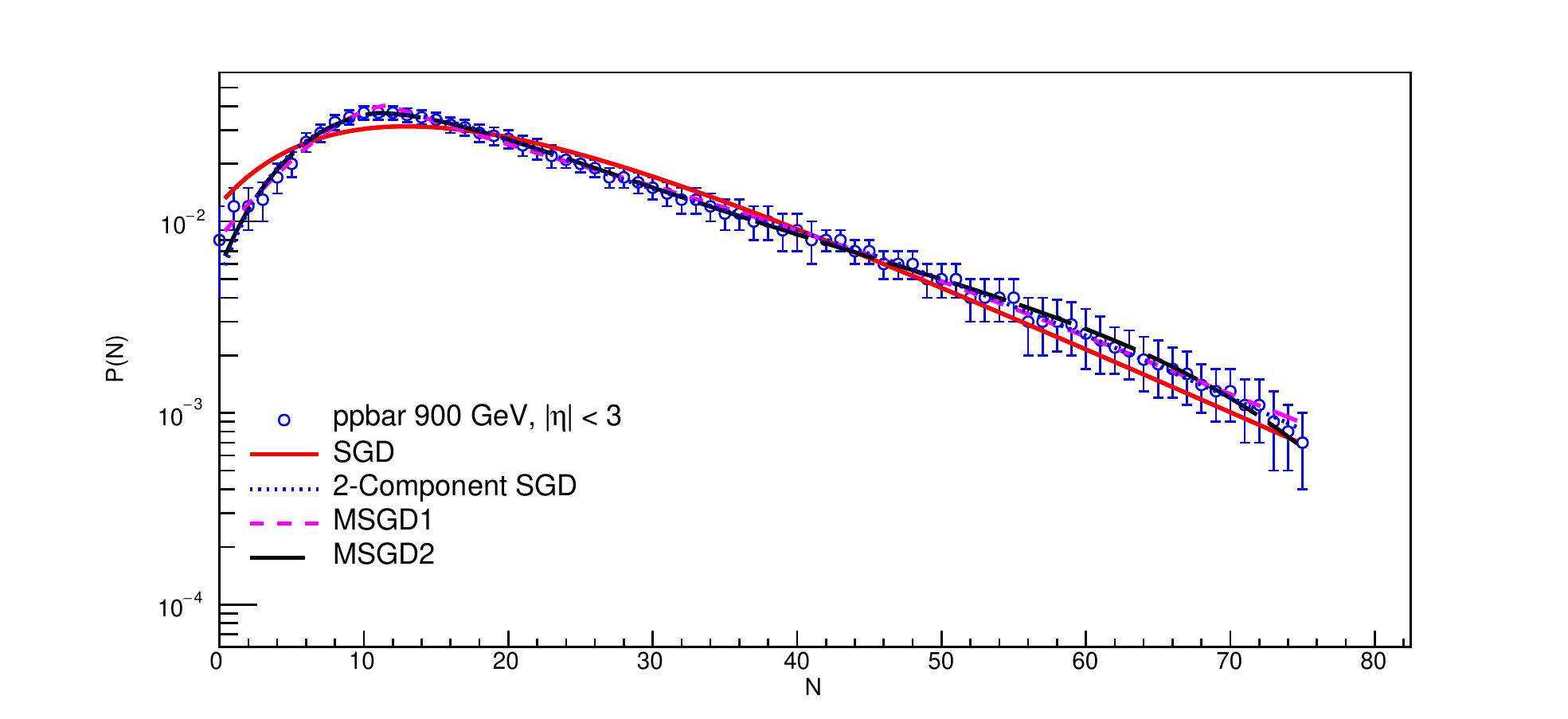}
\includegraphics[width=4.8 in,height =2.43 in]{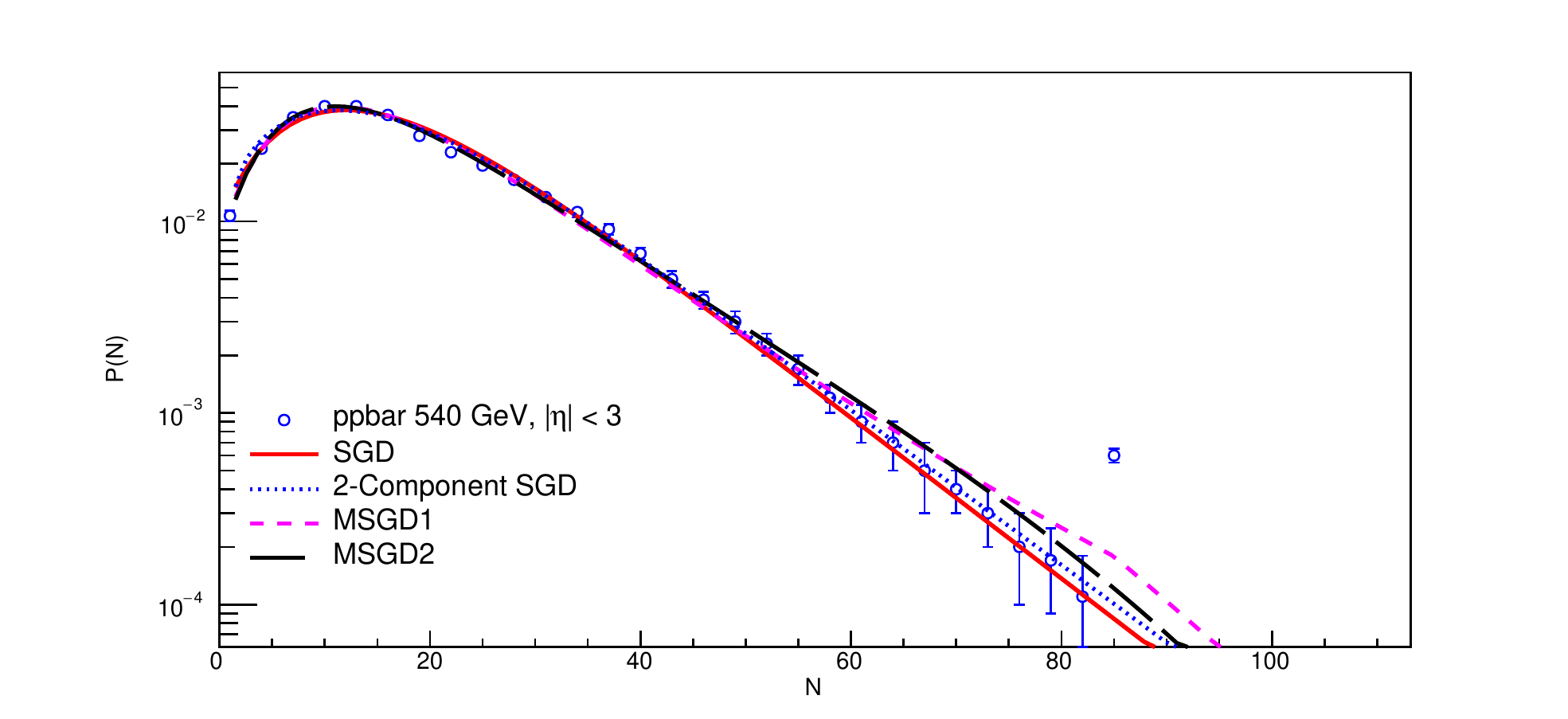}
\includegraphics[width=4.8 in,height =2.43 in]{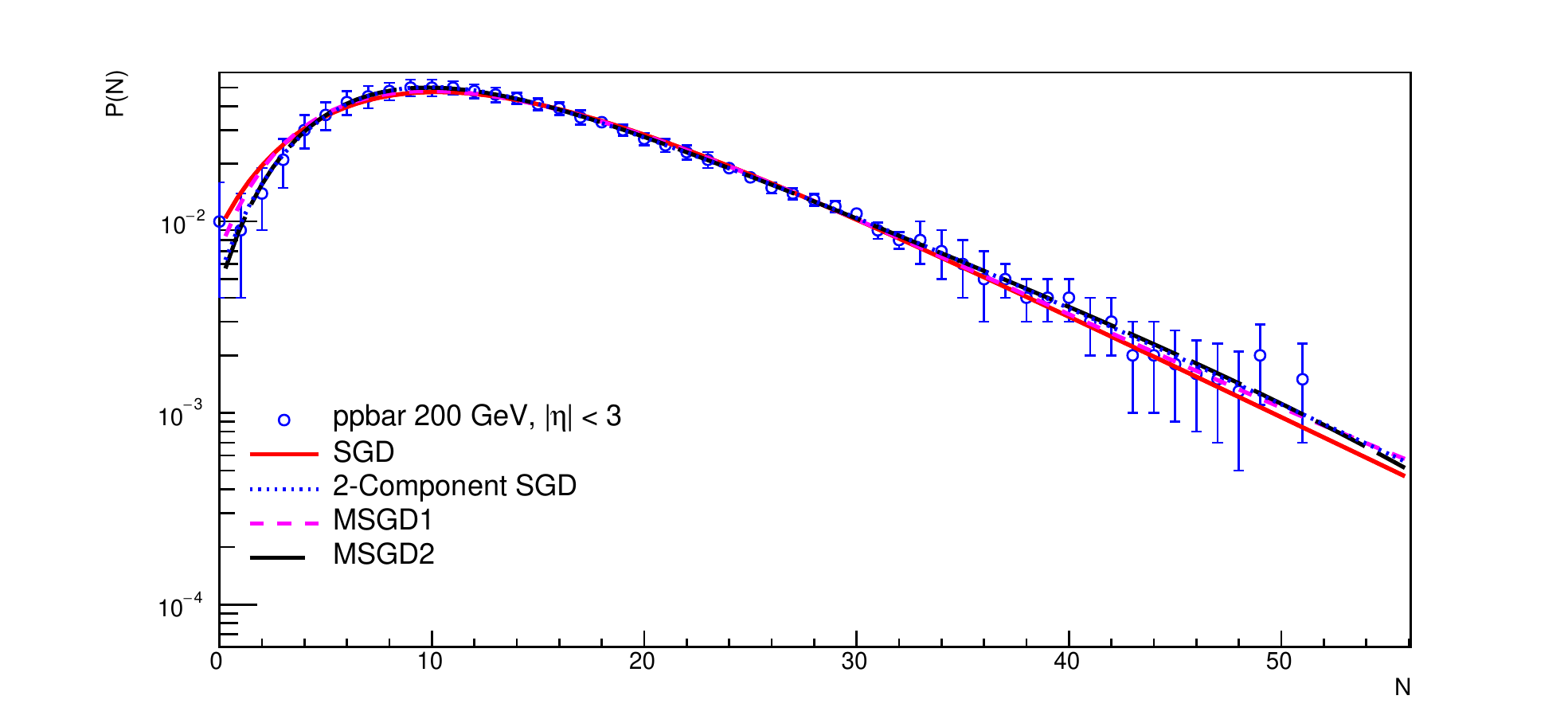}
\caption{SGD, 2-component SGD, MSGD1 and MSGD2 distributions for $p\bar{p}$ collisions at different $\sqrt{s}$ obtained by the UA5 experiment for $|\eta| <3.0$}
\end{figure}  

\begin{table*}[t]
\small
\begin{tabular}{|c|c|c|c|c|c|c|c|}
\hline
Energy   & $Rapidity   $ & $  SGD  $ &  2-component &  $   MSGD1   $ &  $  MSGD2  $ \\ 
$(GeV)$  &  interval           &           &  $SGD$      &                &              \\ \hline           
  & $|\eta|<$ & $\chi^{2}/ndf$ &  $\chi^{2}/ndf$ &  $\chi^{2}/ndf$ &  $\chi^{2}/ndf$ \\ \hline
          
 200 & 0.5 &     11.66 / 11 &      0.43 / 8 &      5.38 / 9 &       0.63 / 8 \\ \hline 
 200 & 1.5 &      9.11 / 29 &      8.79 / 26 &      8.89 / 27 &       9.82 / 26 \\ \hline 
 200 & 3.0 &     12.62 / 48 &      5.23 / 45 &      9.69 / 46 &       5.69 / 45 \\ \hline 
 200 & 5.0 &     35.33 / 52 &      4.40 / 49 &     11.57 / 50 &      34.19 / 49 \\ \hline 
 200 & full &      3.96 / 25 &      2.21 / 22 &      3.50 / 23 &      17.68 / 22 \\ \hline 
 
 540 & 0.5 &     26.90 / 20 &     21.33 / 17 &     19.92 / 18 &      20.53 / 17 \\ \hline 
 540 & 1.5 &     17.22 / 26 &     10.30 / 23 &     15.20 / 24 &       8.20 / 23 \\ \hline 
 540 & 3.0 &    176.38 / 28 &    147.68 / 25 &    130.11 / 26 &     124.13 / 25 \\ \hline 
 540 & 5.0 &     69.33 / 33 &     26.12 / 30 &     54.43 / 31 &      35.54 / 30 \\ \hline 
 540 & full &     59.83 / 49 &     59.83 / 46 &     56.21 / 47 &      34.60 / 46 \\ \hline 
  
 900 & 0.5 &     10.16 / 20 &      5.02 / 17 &      4.73 / 18 &      13.33 / 17 \\ \hline 
 900 & 1.5 &     35.85 / 46 &      3.86 / 43 &      6.12 / 44 &      15.53 / 43 \\ \hline 
 900 & 3.0 &     63.90 / 72 &      6.97 / 69 &      8.57 / 70 &       8.57 / 69 \\ \hline 
 900 & 5.0 &     89.95 / 95 &     89.95 / 92 &     34.81 / 93 &      25.06 / 92 \\ \hline
 900 & full &     67.16 / 47 &     11.23 / 44 &     15.67 / 45 &      13.69 / 44 \\ \hline  

\end{tabular}
\caption{$\chi^{2}/ndf$ for charged multiplicity distribution fitted with Shifted Gompertz, 2-component modified Shifted Gompertz, MSGD1 and MSGD2 distributions for different rapidity windows in $p\bar{p}$ collisions.}
\end{table*}

In figure~3 and~4, we show the ratio plots for multiplicity dependence of the ratio $R = P_{data}(N)/P_{fit}(N)$ for the $pp$ data shown in figure~1 obtained from the fits SGD and MSGD2.~Figure~5 shows similar ratio plots for multiplicity dependence of the ratio for the $p\overline{p}$ data shown in figure~2.~As can be seen from figures~3 and 5, there are systematic deviations from the fits of SGD from the data at low and high multiplicities.~The deviations get enhanced with increasing energy and high multiplicity values, as can also be observed in figures~1 and 2.~In addition, a structure at smaller multiplicities can also be observed.~In order to understand this structure, the modified forms of SGD have been introduced as MSGD1 and MSGD2 in equations~(7,8).~The ratio $R$ calculated with MSGD2, becomes closer to unity in all the cases, though the deviations are still present.~The possibility of retrieving some additional information from experimental multiplicity distribution, the recurrence relation given in equation (9) is used to calculate the coefficients $C_{j}$.~In some cases, the 2-component SGD fits exceptionally well leading to the $R$ value around unity, as shown in figure~4.
 
\begin{figure}[ht]
\includegraphics[width=4.8in,height =2.43 in]{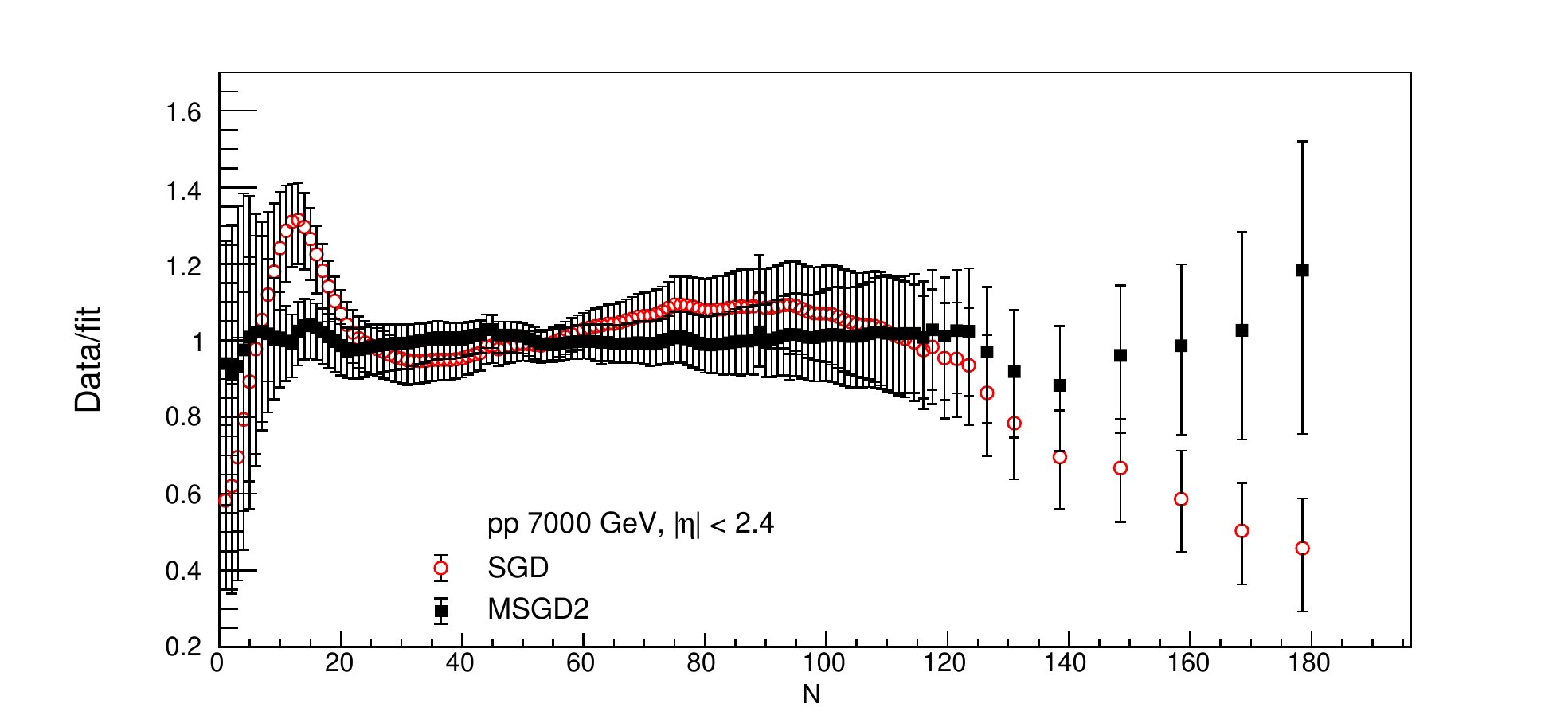}
\includegraphics[width=4.8 in,height =2.43 in]{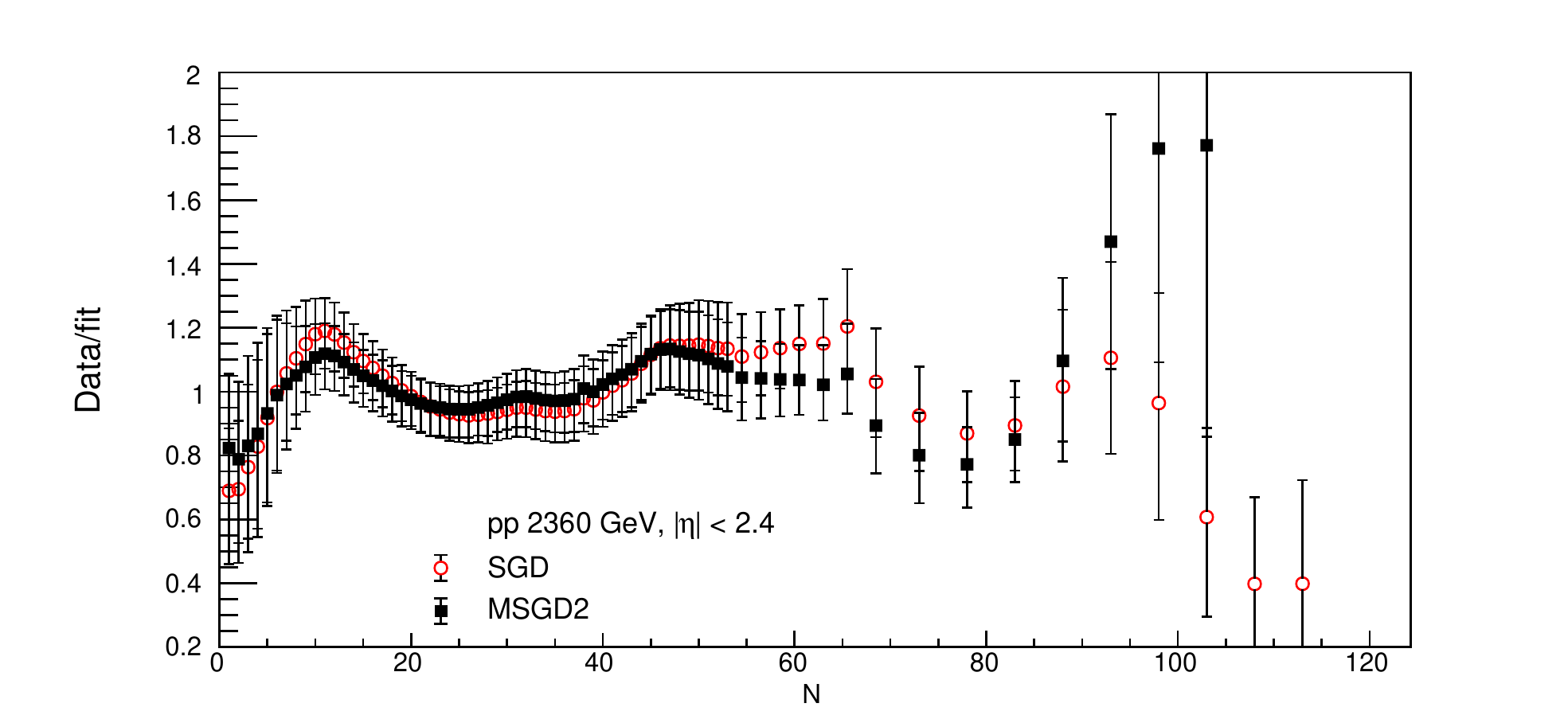}
\includegraphics[width=4.8 in,height =2.43 in]{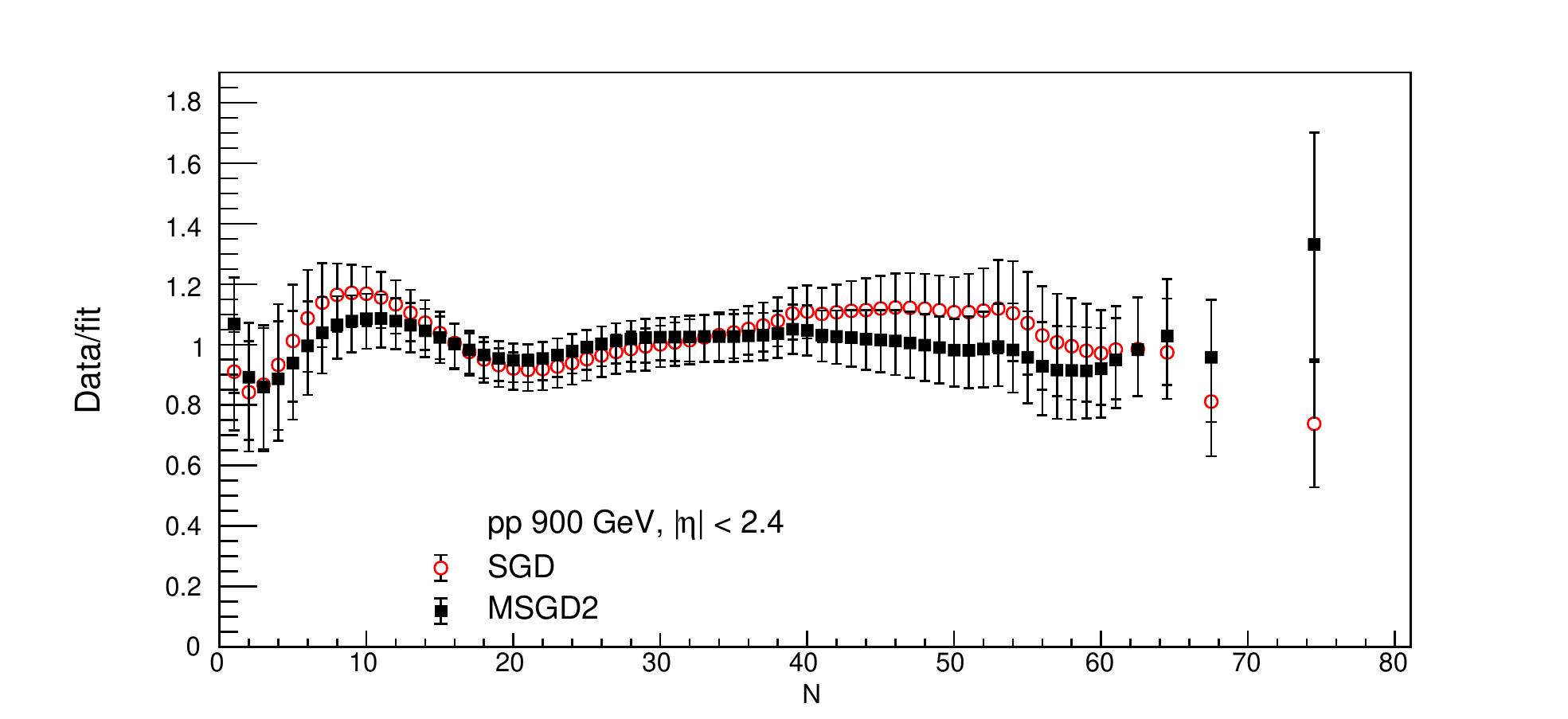}
\caption{Ratio $R = P_{data}(N)/P_{SGD}(N)$ for the $pp$ data shown in figure~1 (red circles) and of the corrected ratio $R = P_{data}(N)/P_{MSGD2}(N)$ (black squares).}
\end{figure}

\begin{figure}[ht]

\includegraphics[width=4.8 in,height =2.43 in]{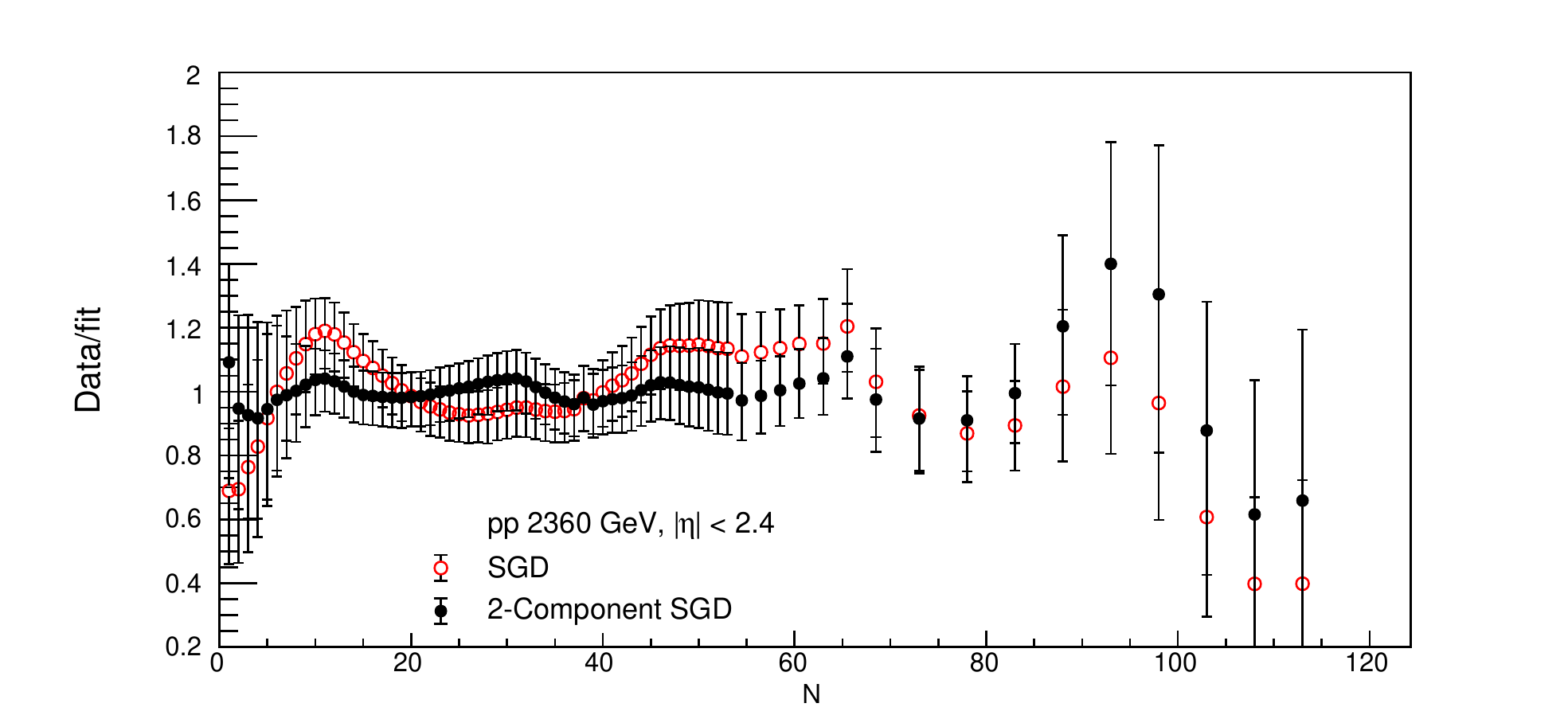}
\includegraphics[width=4.8 in,height =2.43 in]{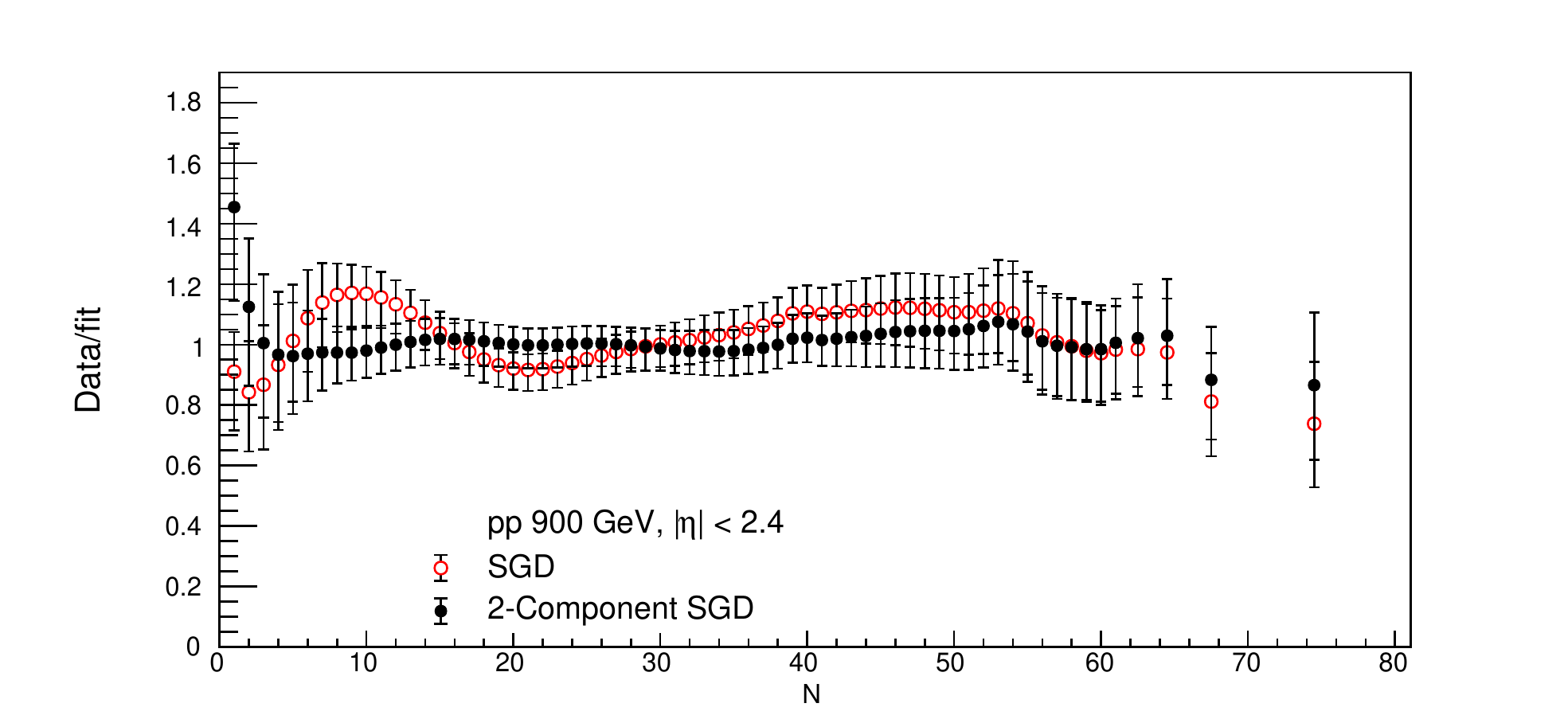}
\caption{Ratio $R = P_{data}(N)/P_{SGD}(N)$ for the $pp$ data shown in figure~1 (red circles) and of the corrected ratio $R = P_{data}(N)/P_{2-component SGD}(N)$ (black circles).}
\end{figure}

\begin{figure}[ht]
\includegraphics[width=4.8 in, height =2.43 in]{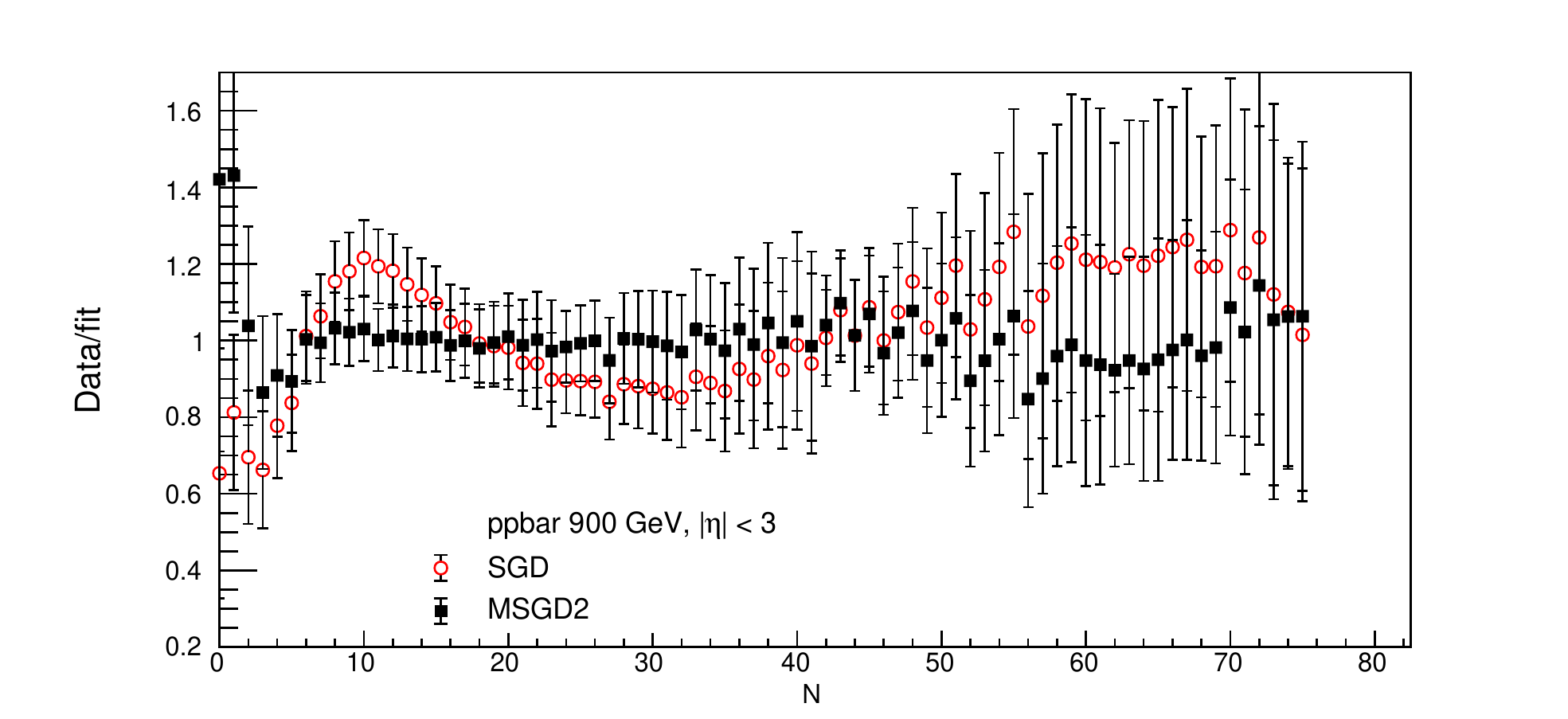}
\includegraphics[width=4.8 in, height =2.3 in]{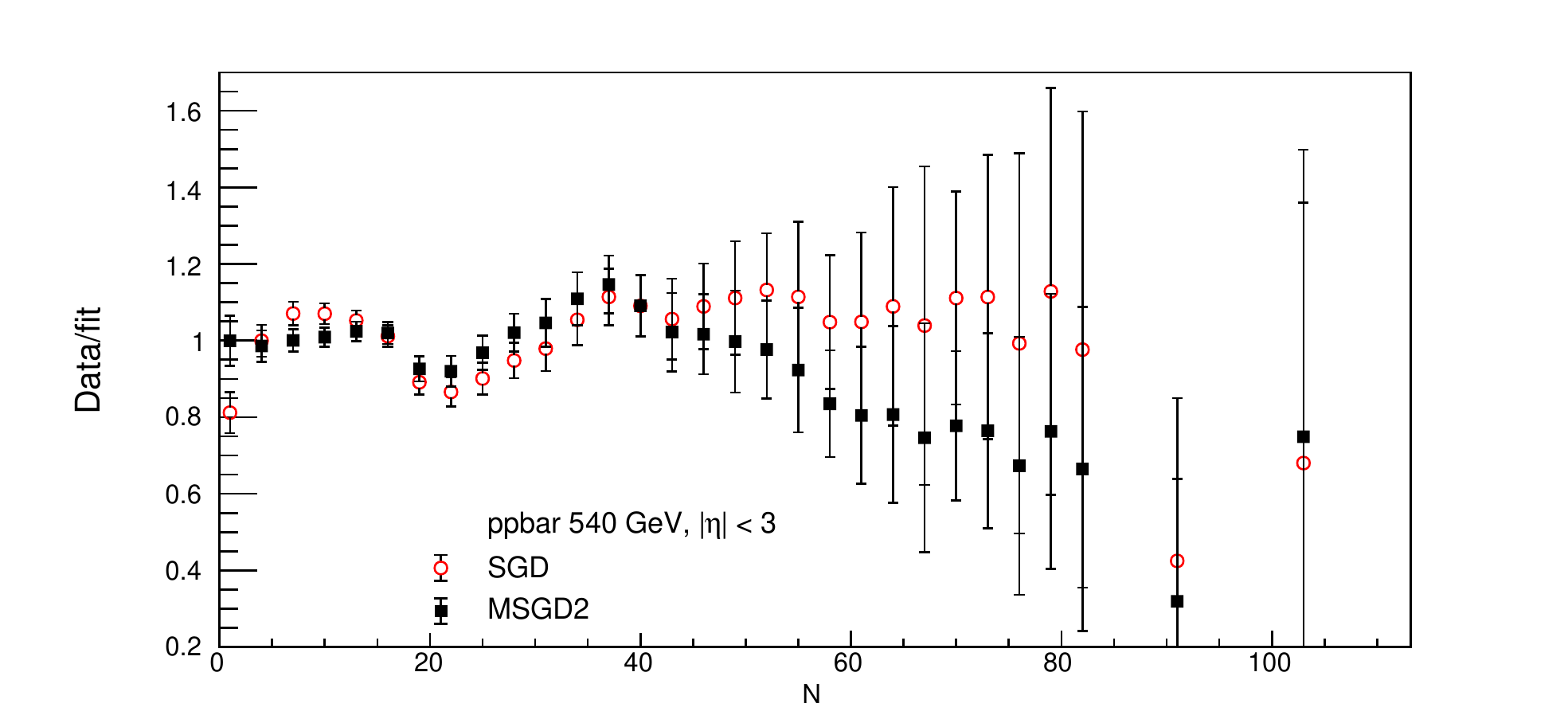}
\includegraphics[width=4.8 in, height =2.43 in]{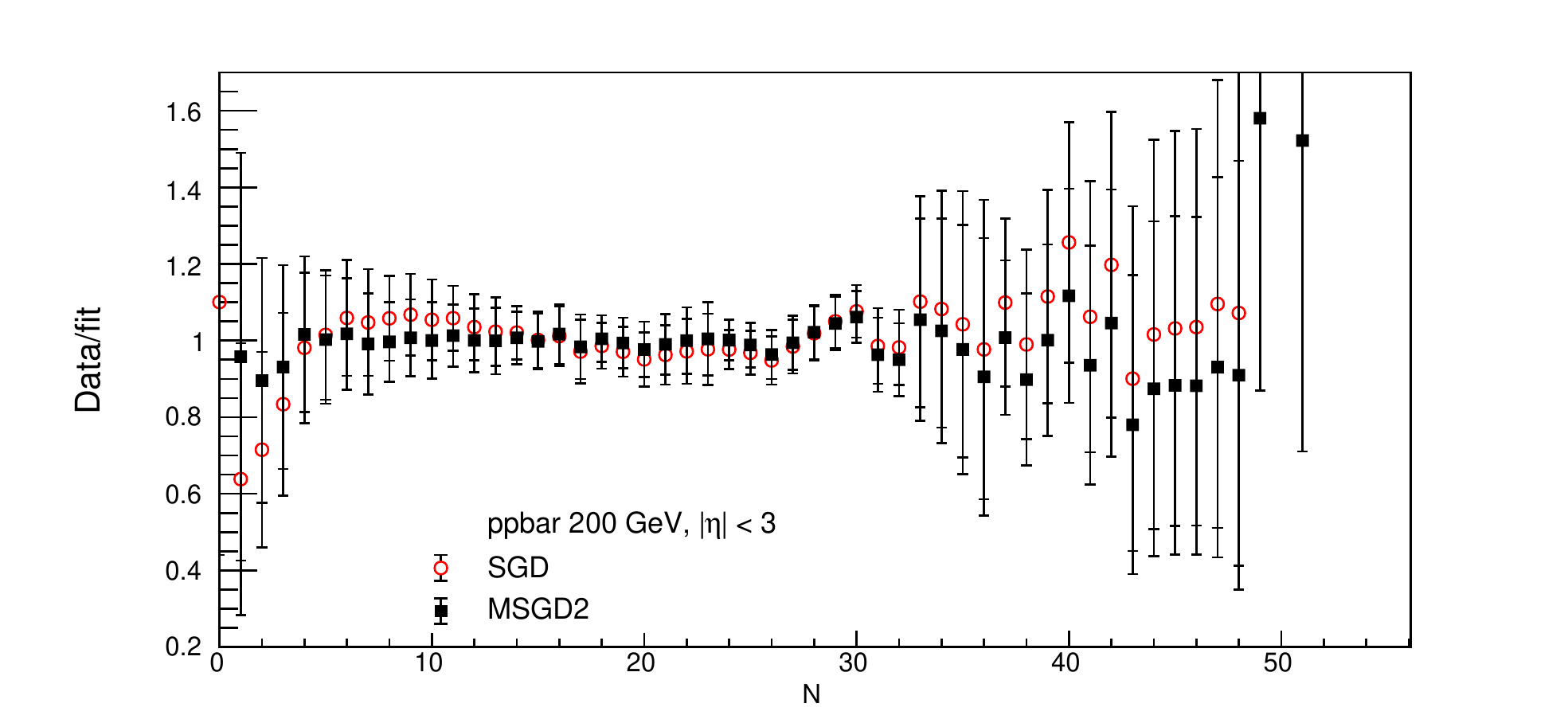}
\caption{Ratio $R = P_{data}(N)/P_{SGD}(N)$ for the $p\overline{p}$ data shown in figure~2 (red circles) and of the corrected ratio $R = P_{data}(N)/P_{MSGD2}(N)$ (black squares)}
\end{figure}

The coefficients $C_{j}$ are calculated for the $pp$ and $p\overline{p}$ data at different centre-of-mass energies and for various pseudo-rapidity windows.~Figure~6 shows $C_{j}$ for $pp$ data in $|\eta| <$ 2.4 and for the $p\overline{p}$ data for $|\eta|<$ 3.0.~The figures show a very distinct oscillatory behaviour in both the cases.~For the case of $pp$ interactions, oscillations occur with amplitude decreasing with the rank $j$ at all energies.~It shows that the effect of an increase in centre-of-mass collision energy has minimal effect on the amplitude and the period of the resulting oscillations.~However for the $p\overline{p}$ interactions, the trend is reversed, with the amplitude of oscillations growing with the rank $j$ and decrease in collision energy.~This intriguing property has also been observed by Rybczy$\acute{•}{n}$ski et al \cite{Ryb}.~The way the $C_j$ oscillates between $pp$ and $p\overline{p}$ collisions is clearly different and may be a characteristic of matter-antimatter collision.~Abramovsky and Radchenko in their paper {\cite{Abram} have described the particle production in inelastic collisions in terms of quark-and-gluon strings.~They have described the multiplicity distributions in terms of 2-NBD and 3-NBD and have shown how the $pp$ and $p\overline{p}$ collisions are fundamentally different, and which may lead to the observed differences.~In another interesting study by H.W. Ang et al \cite{Ang}, similar differences have been observed in $p\overline{p}$ (UA5) and $pp$ (ALICE) data.

~Figure~7 shows the coefficients $C_{j}$ calculated for the $pp$ collision data at 7000~GeV c.m. energy and for $p\overline{p}$ collisions at 200~GeV, for different pseudo-rapidity windows.~They all show the distinct oscillatory behaviour with amplitude increasing with pseudorapidity window for both $pp$ and $p\overline{p}$ collisions.~It is also observed that the oscillations die out with increasing rank $j$ for all $|\eta|$ bins for $pp$ collisions, whereas for $p\overline{p}$ collisions, the oscillations grow stronger with rank $j$ with increasing $|\eta|$ bin size and somewhat random only in $|\eta|<$ 5.0 bin.~Similar observations are also observed by Rybczy$\acute{•}{n}$ski et al \cite{Ryb} in the CMS and ALICE data \cite{CMS,ALICE}.~In figures~6-7, the errors on the data points are not shown for the reason that the error bars intermingle and blur the figures.

\begin{figure}[ht]
\includegraphics[width=4.8 in,height =2.8in]{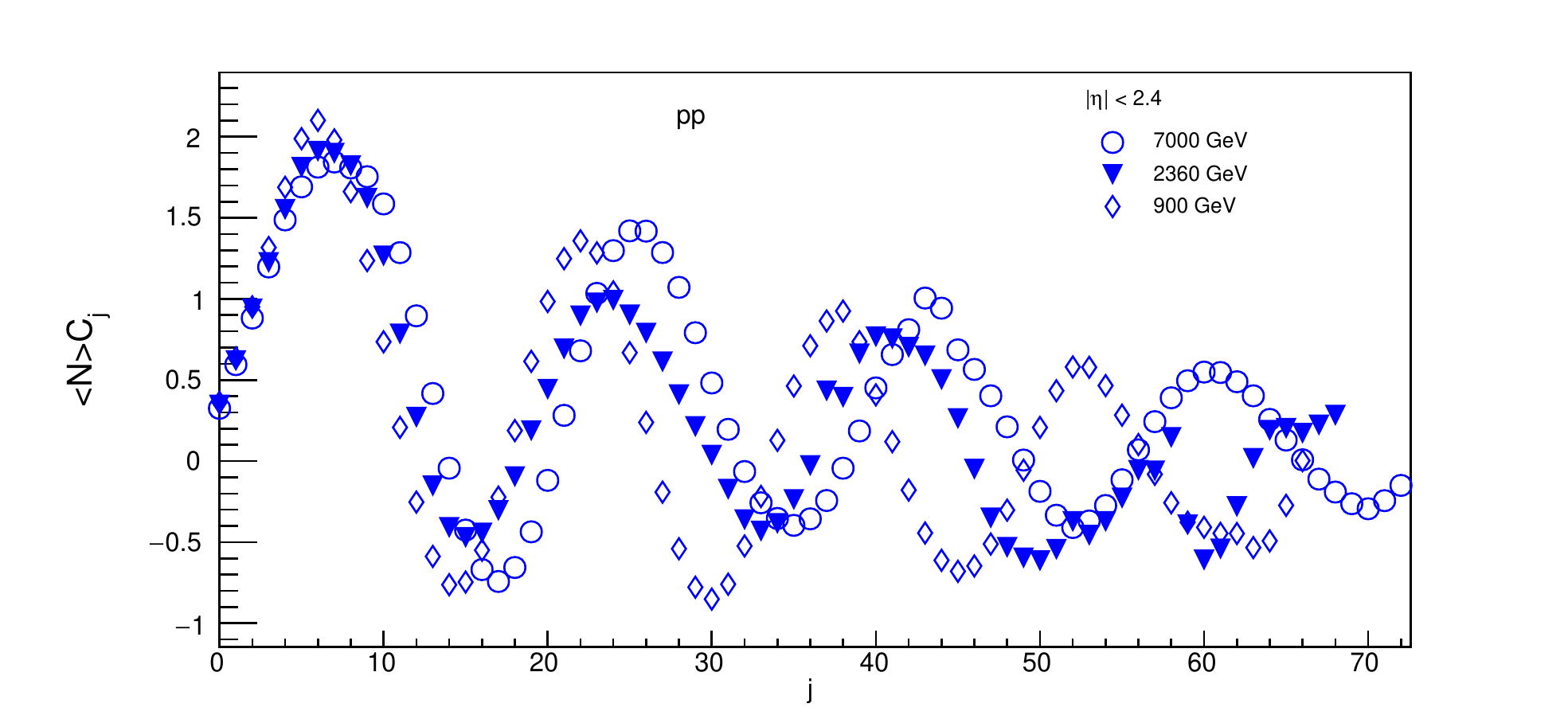}
\includegraphics[width=4.8 in,height =2.8in]{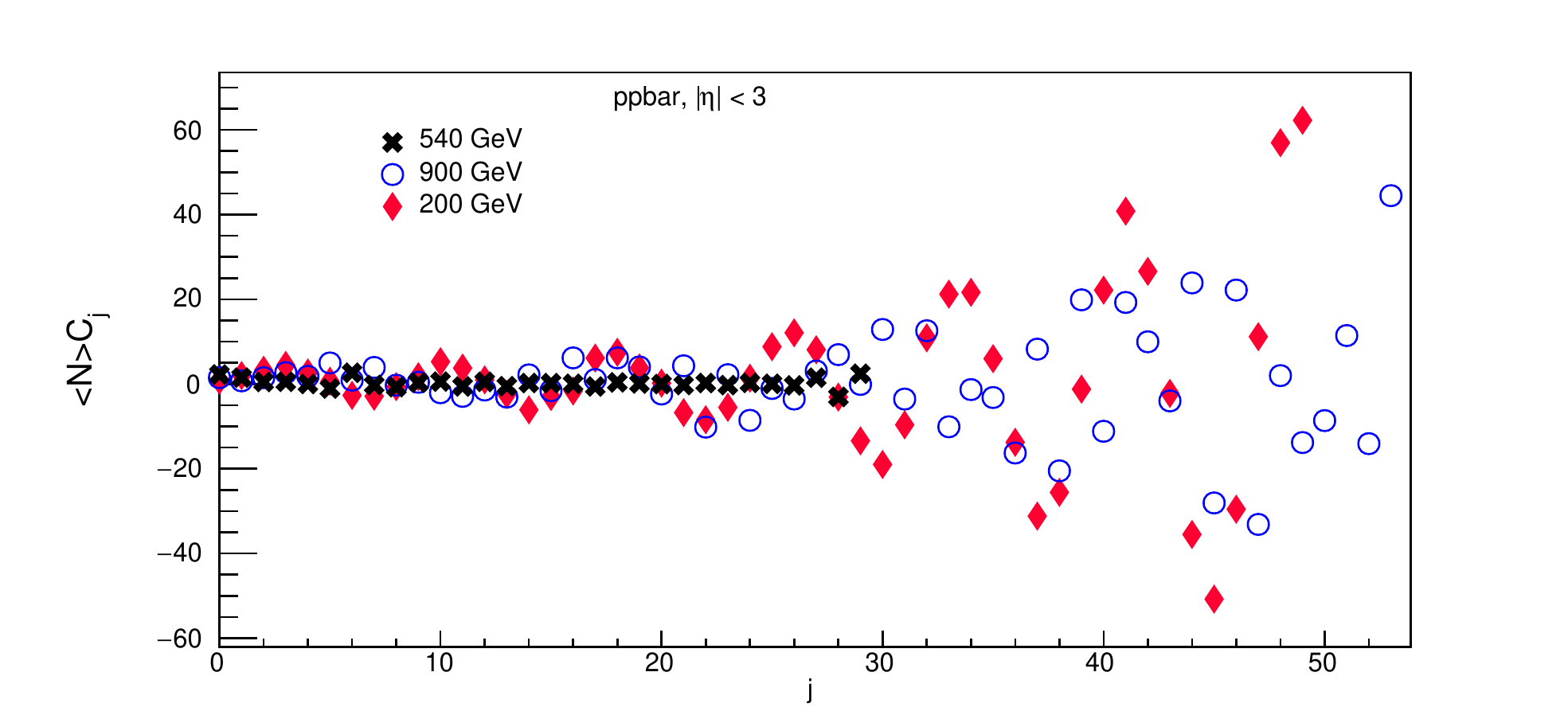}
\caption{Coefficients $C_j$ obtained from i) the CMS data of $pp$ collisions at different energies for one pseudorapidity window $|\eta|<2.4$(top), ii) the UA5 data of $\bar{p}p$ collisions at different energies for pseudorapidity window $|\eta|<3$ (bottom).} 
\end{figure}

\begin{figure}[ht]
\includegraphics[width=4.8 in,height =2.8 in]{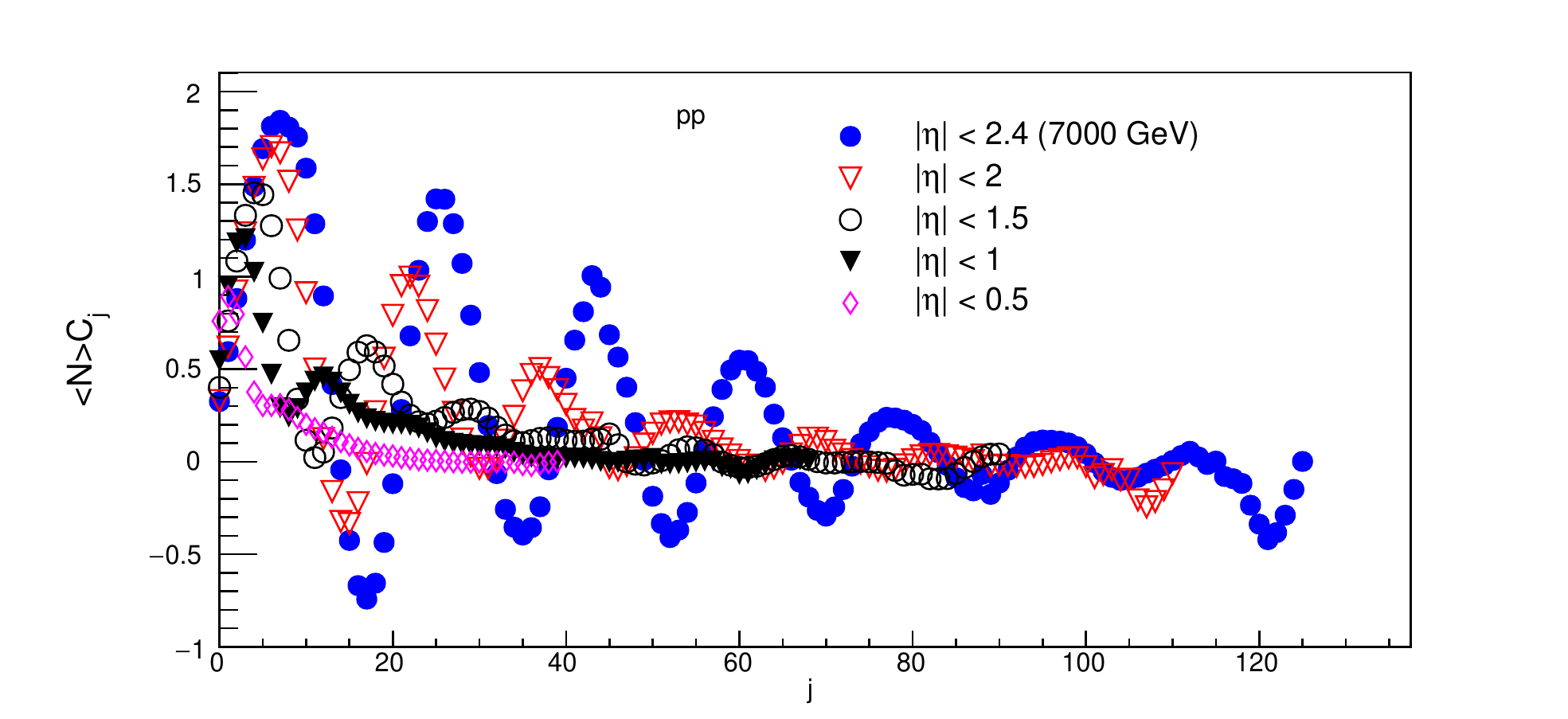}
\includegraphics[width=4.8 in,height =2.8 in]{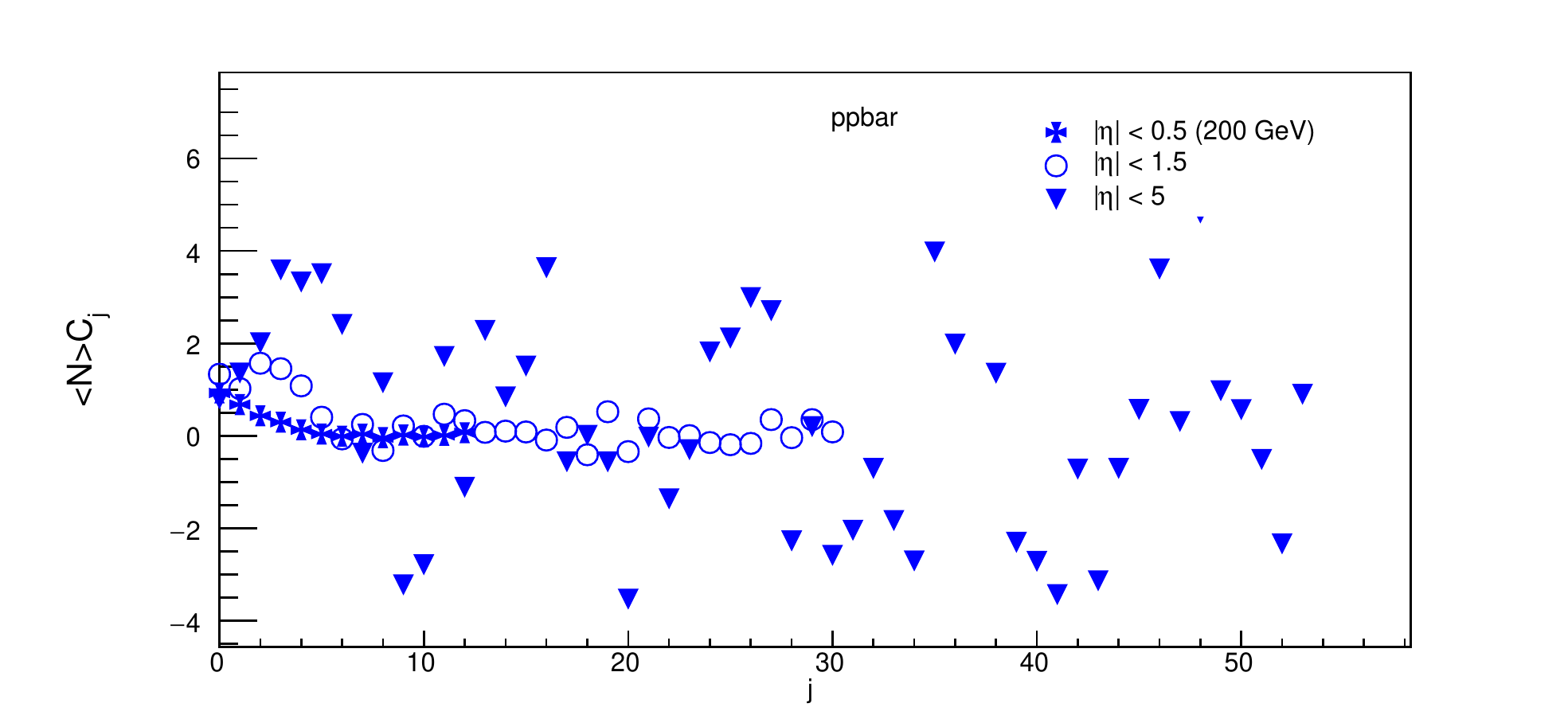}

\caption{Coefficients $C_j$ obtained from the CMS data of $pp$ collisions at $\sqrt{s}$ = 7000~GeV for different pseudorapidity windows (top) and the UA5 data of $\bar{p}p$ collisions at $\sqrt{s}$ = 200~GeV in different pseudorapidity windows (bottom).}
\end{figure}

The coefficients $C_j$ are evaluated by fitting the SGD, 2-component SGD, MSGD1 and MSGD2 distributions to the data.~The variation of these coefficients with $j$ are shown in figure~8 for $pp$ data in one pseudo-rapidity window for different energies.~We find that the $C_j$ evaluated from the SGD fit do not show this oscillatory behaviour, as compared with the data.~However, with the 2-component fit, they start showing the oscillatory pattern, which further gets enhanced with MSGD1 and MSGD2 fits, following the data closely.~In case of MSGD2 fits, coefficients $C_{j}$ follow almost exactly the oscillatory behaviour of the $C_{j}$  obtained directly from the data at $\sqrt{s}$ = 7000~GeV.~While for $\sqrt{s}$ = 2360~GeV, it is MSGD1 and for $\sqrt{s}$ = 900~GeV, the 2-component SGD follow the experimental values better.~Similar results are obtained by analysing the data for different pseudorapidity windows both of $pp$ and $p\overline{p}$ collisions.~However, we show the results for 7000, 2360 and 900 GeV $pp$ collisions for only $|\eta|<$2.4 and similarly for 900, 540 and 200 GeV $p\overline{p}$ collisions for $|\eta|<$1.5, in figure~9.~It may be observed that none of the fits consistently follow the $p\overline{p}$ data trends.~This is also seen for other $\eta$ windows.~To avoid too many similar figures, we present only the representative figures.

\begin{figure}[ht]
\includegraphics[width=4.8 in,height =2.4 in]{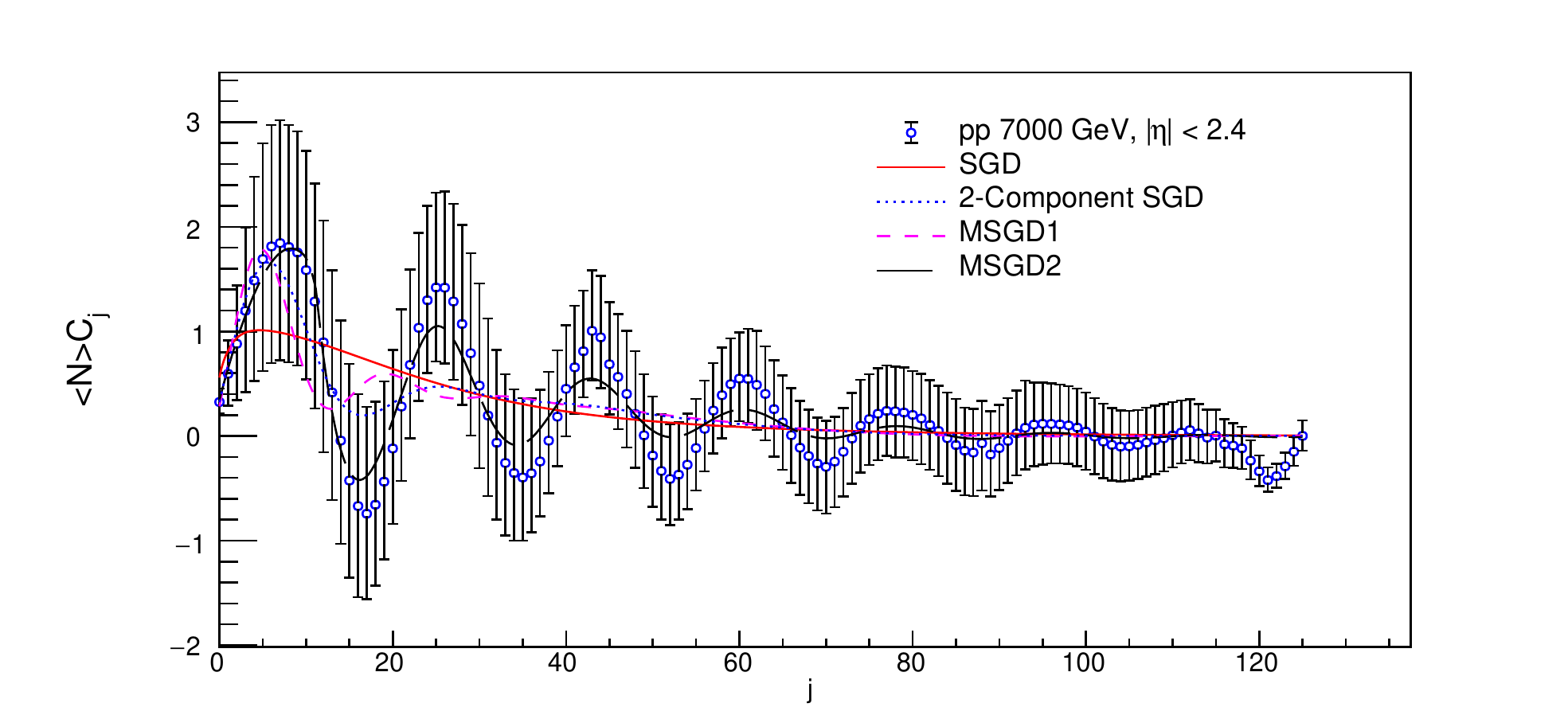}
\includegraphics[width=4.8 in,height =2.4 in]{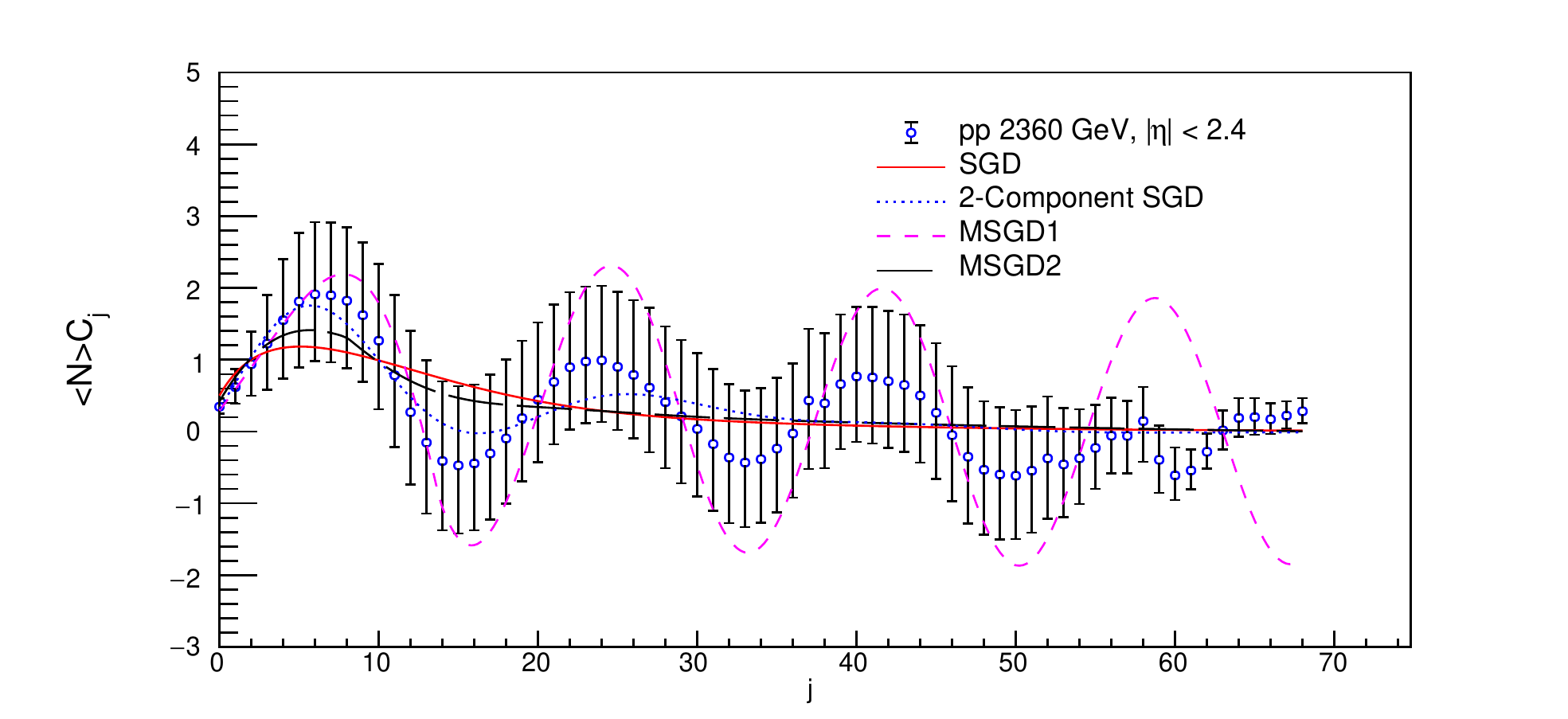}
\includegraphics[width=4.8 in,height =2.4 in]{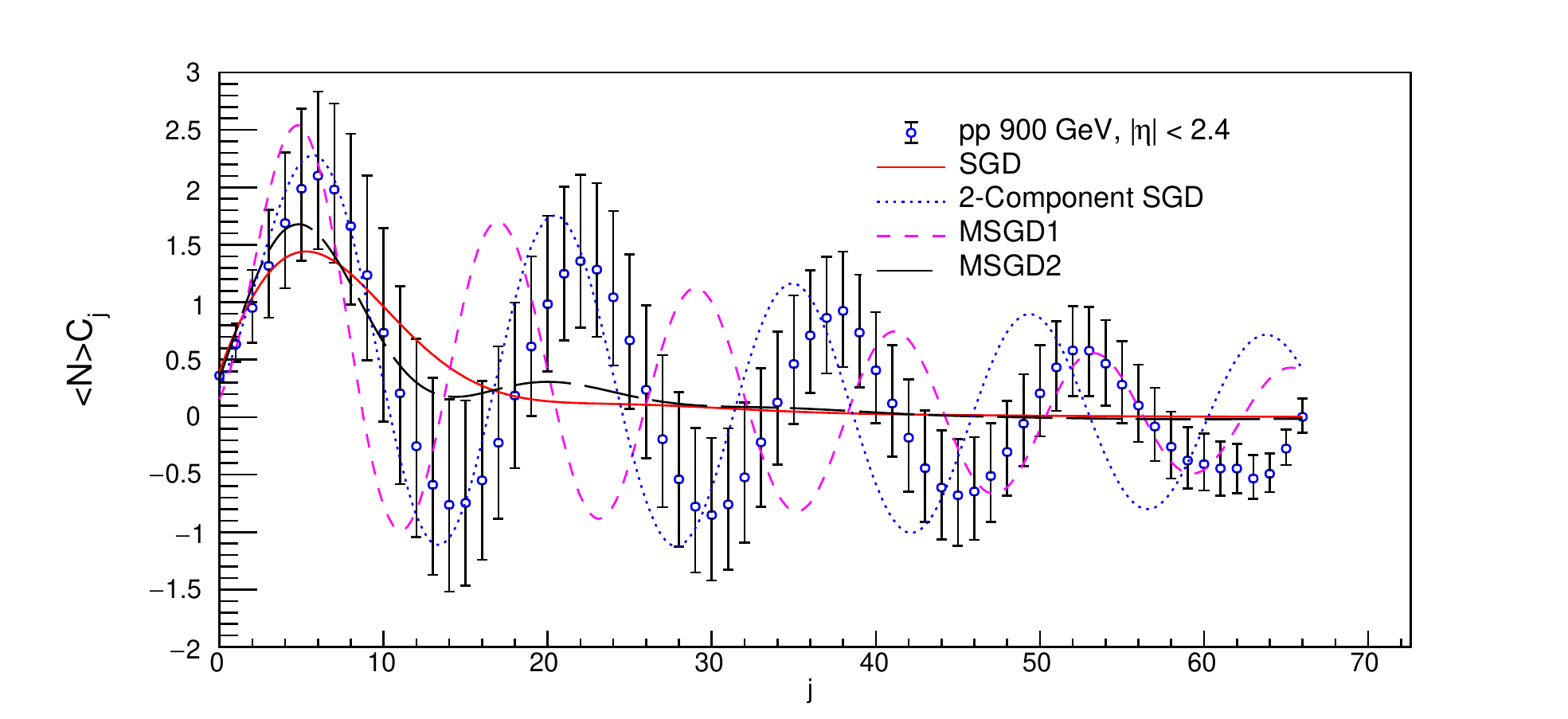}
\caption{Coefficients $C_j$ obtained from the CMS data of $pp$ collisions at different energies in one pseudorapidity window compared with the $C_j$ obtained from SGD, 2-component SGD, MSGD1 and MSGD2 distribution fits to the data}
\end{figure}

\begin{figure}[ht]
\includegraphics[width=4.8 in,height =2.38 in]{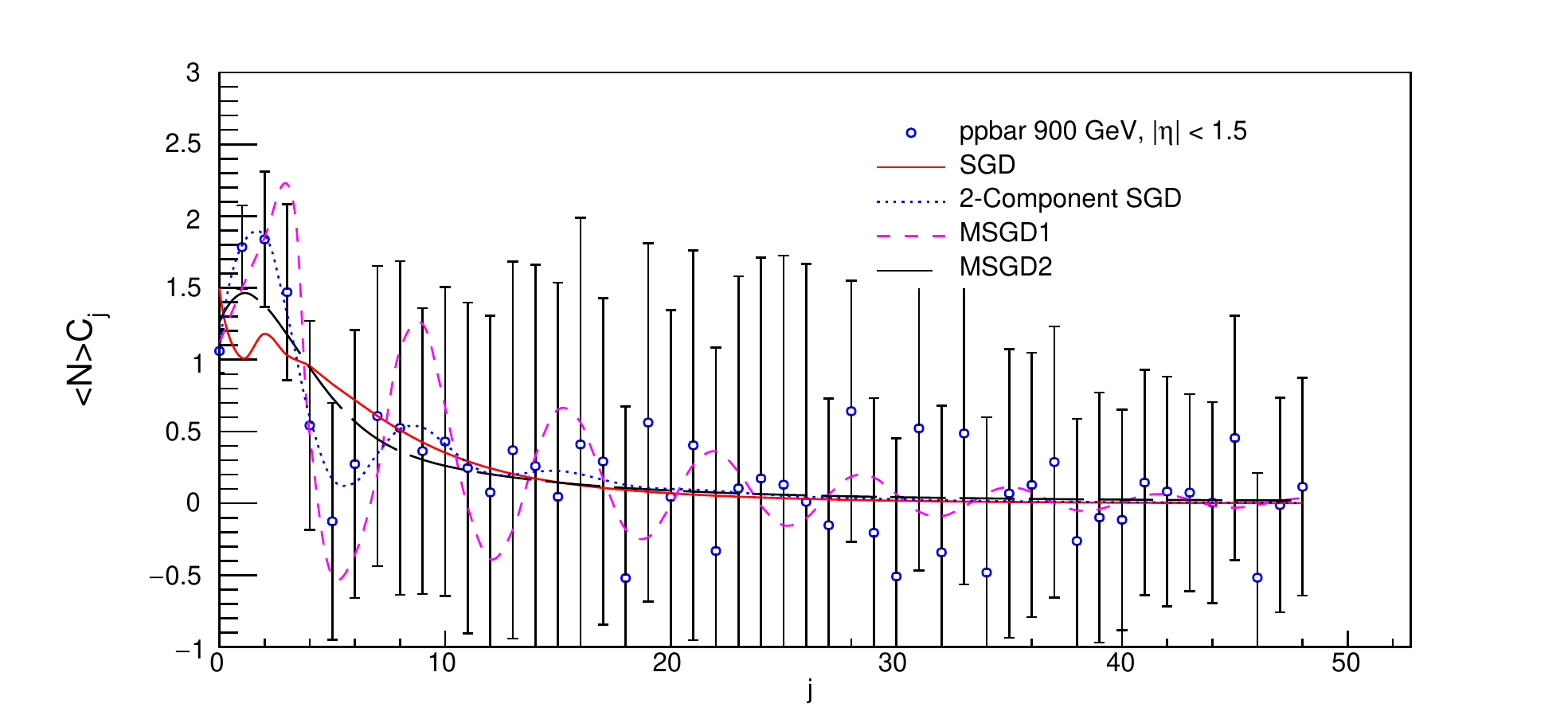}
\includegraphics[width=4.8 in,height =2.38 in]{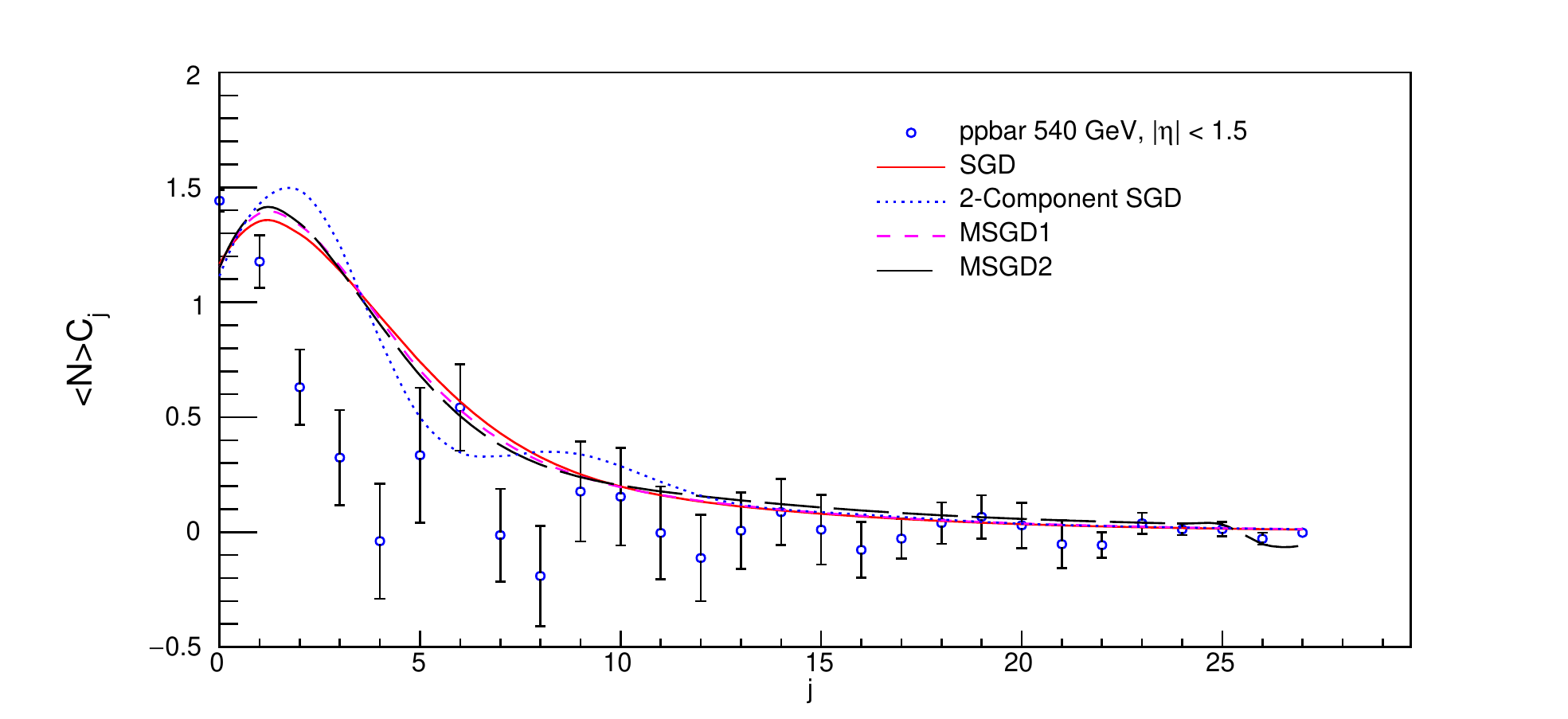}
\includegraphics[width=4.8 in,height =2.38 in]{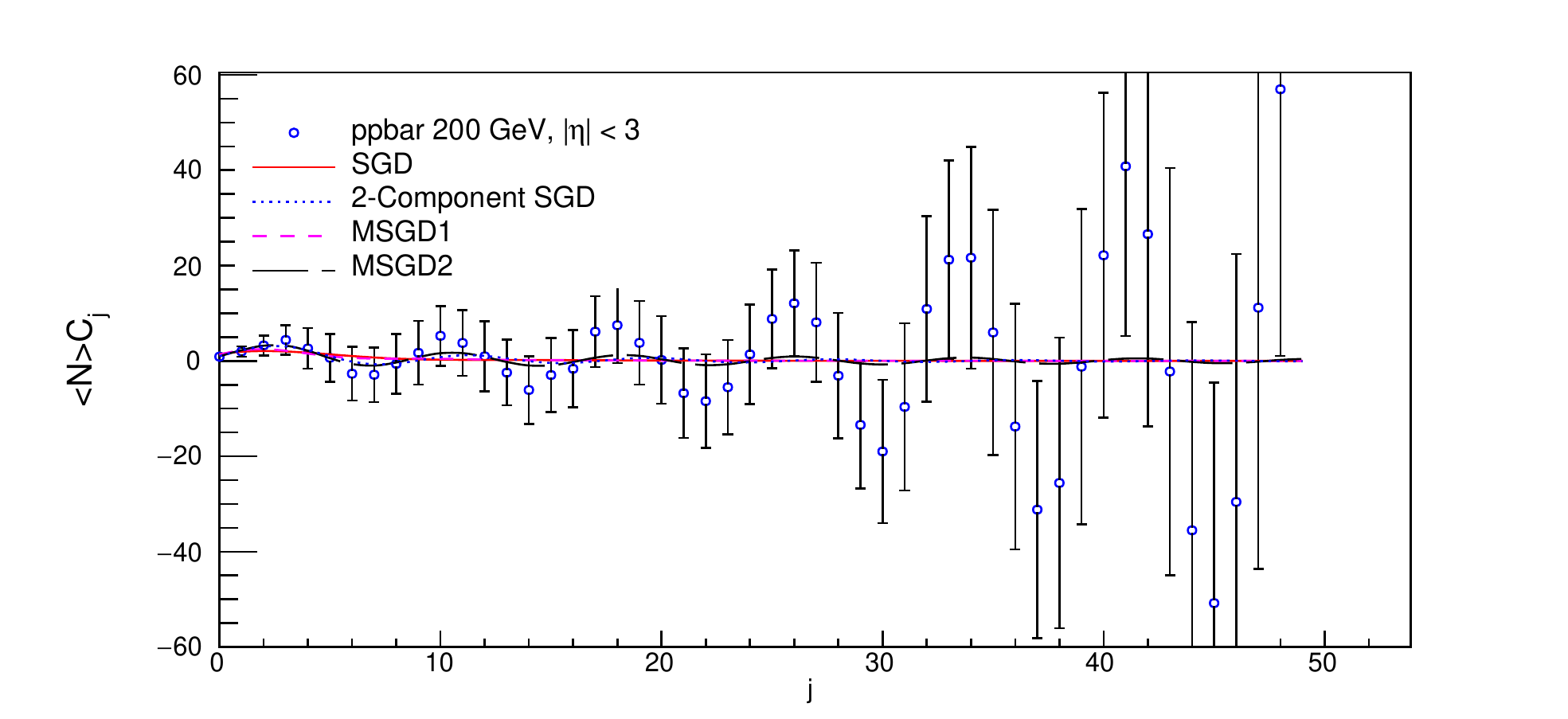}
\caption{Coefficients $C_j$ obtained from the UA5 data of $p\overline{p}$ collisions in one pseudorapidity window compared with the $C_j$ obtained from the SGD, 2-component SGD, MSGD1 and MSGD2 distribution fits to the data at different $\sqrt{s}$.}
\end{figure}

\begin{figure}[ht]
\includegraphics[width=4.8 in,height =2.8 in]{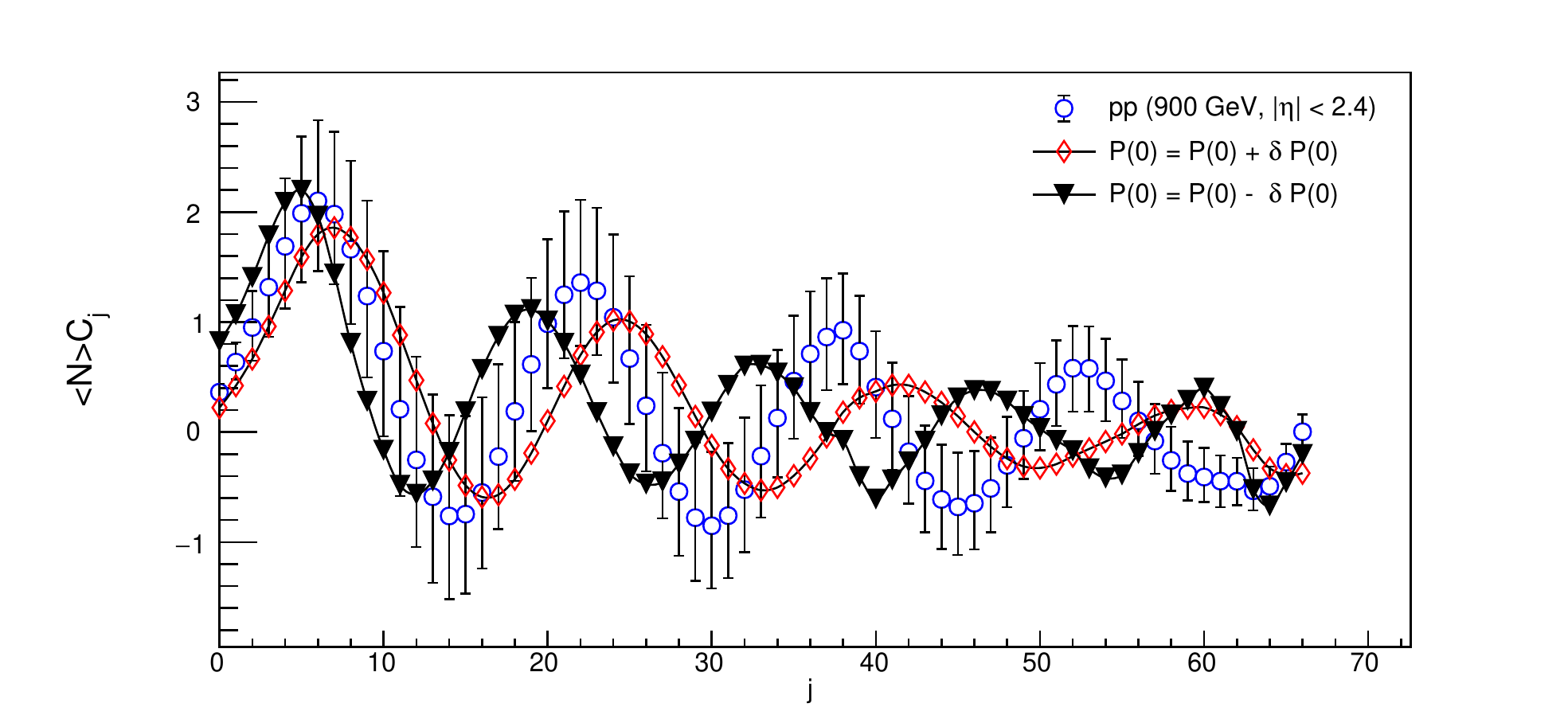} 
\caption{Illustration of oscillatory behaviour of the coefficients $C_j$, using error limits on the probability $P(0)$.}
\end{figure}
\begin{figure}[ht]
\includegraphics[width=4.8 in,height =2.8 in]{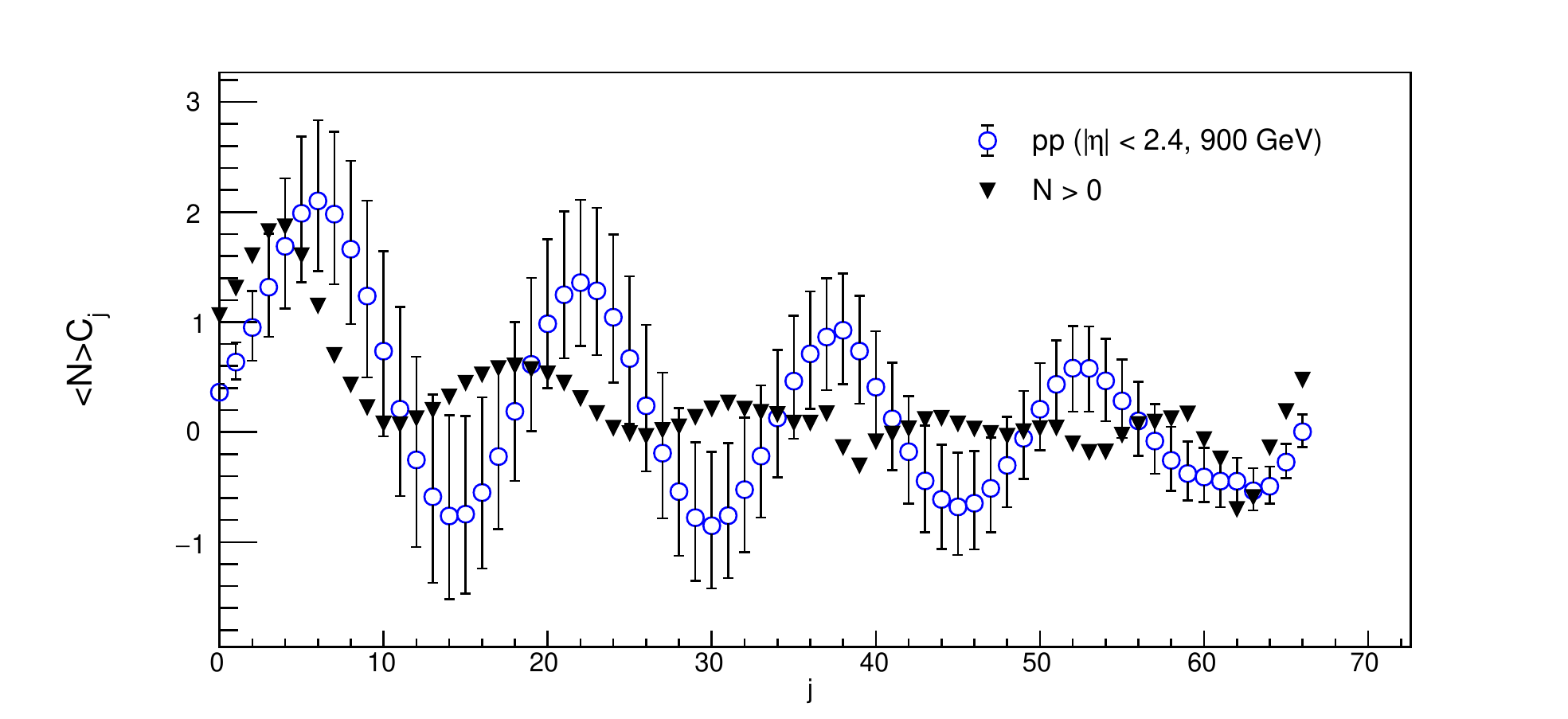} 
\caption{Illustration of oscillatory behaviour of the coefficients $C_j$, removing the probability $P(0)$ and starting with $P(1)$.}
\end{figure}

\begin{figure}[ht]
\includegraphics[width=4.8 in,height =2.48 in]{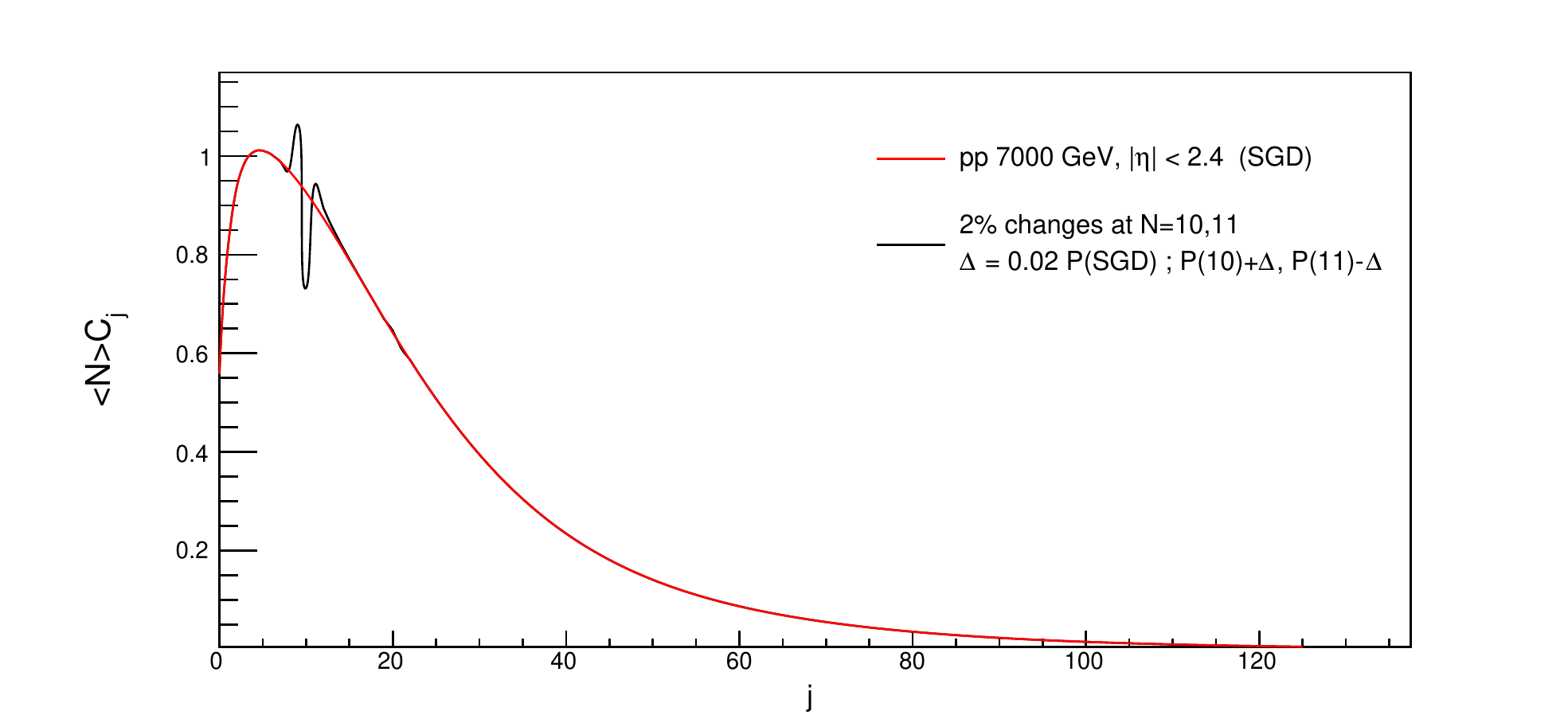}
\includegraphics[width=4.8 in,height =2.48 in]{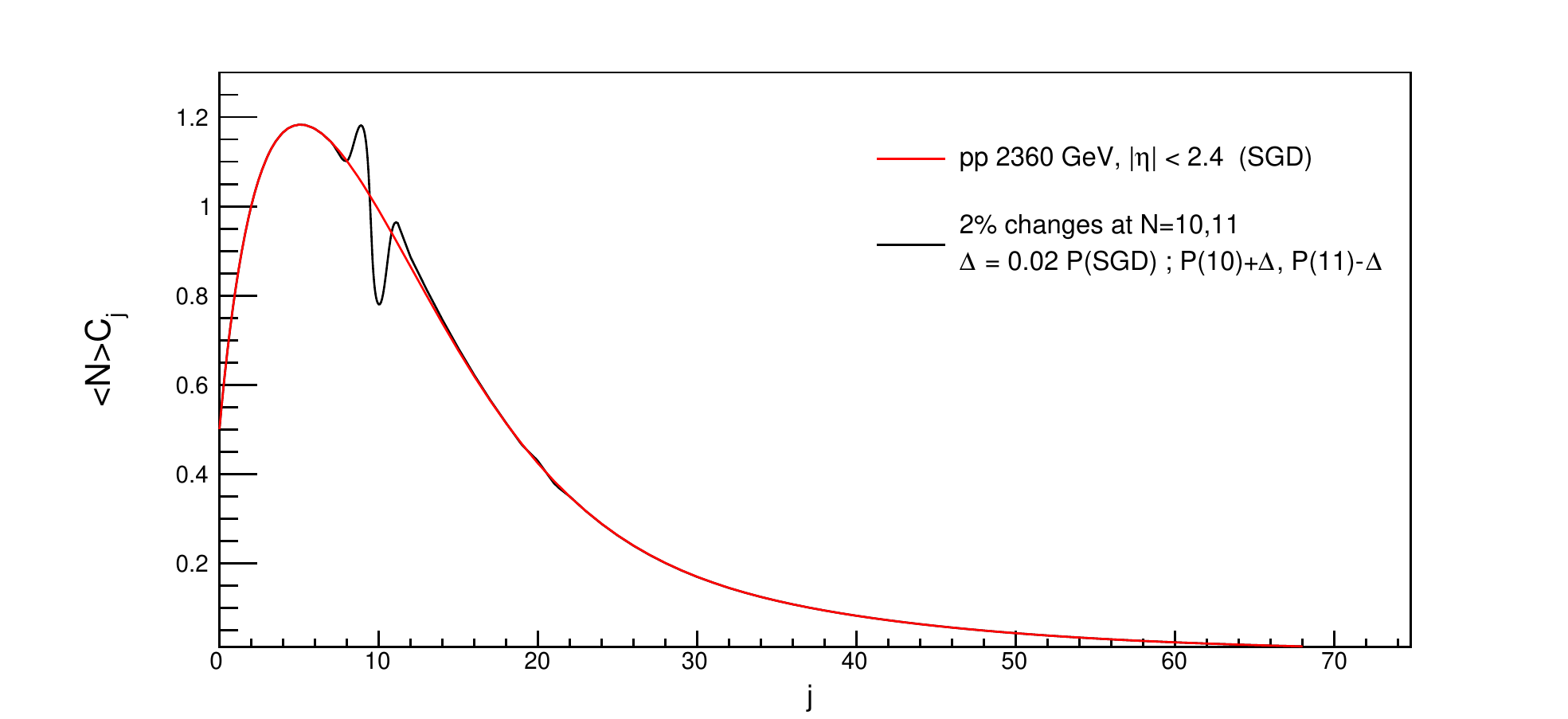}
\includegraphics[width=4.8 in,height =2.48 in]{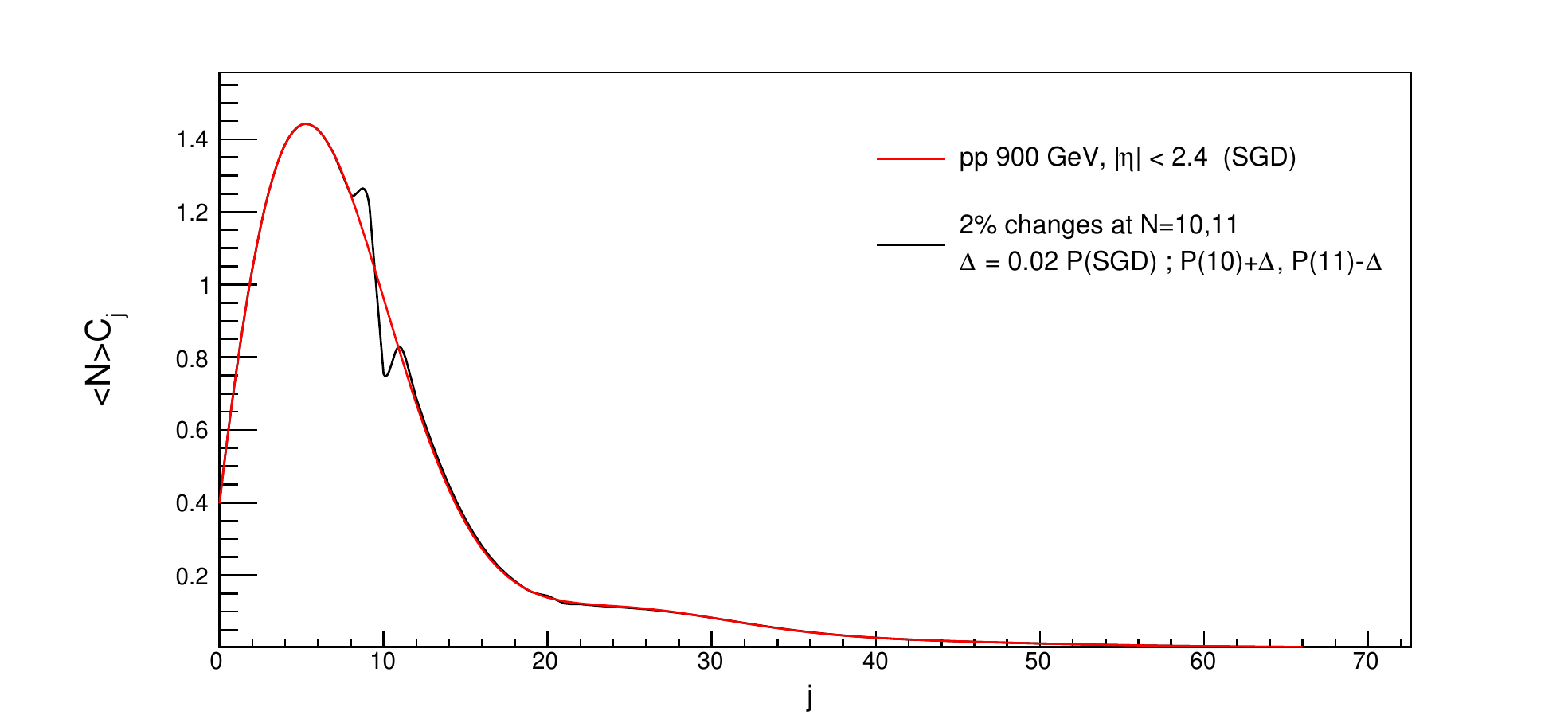}
\caption{Illustration of oscillatory behaviour of the coefficients $C_j$, as described in the text.}
\end{figure}

\begin{figure}[ht]
\includegraphics[width=4.8 in,height =2.48 in]{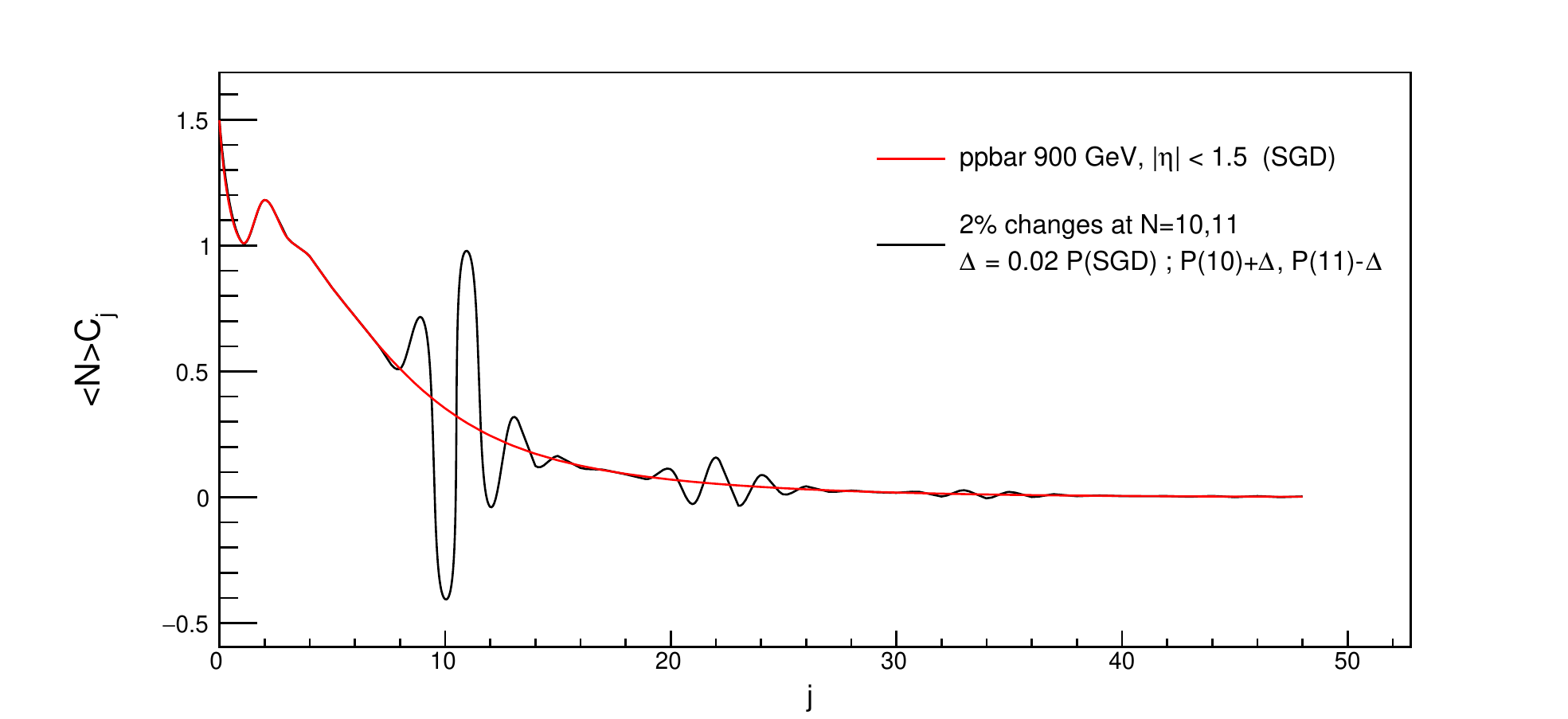}
\includegraphics[width=4.8 in,height =2.48 in]{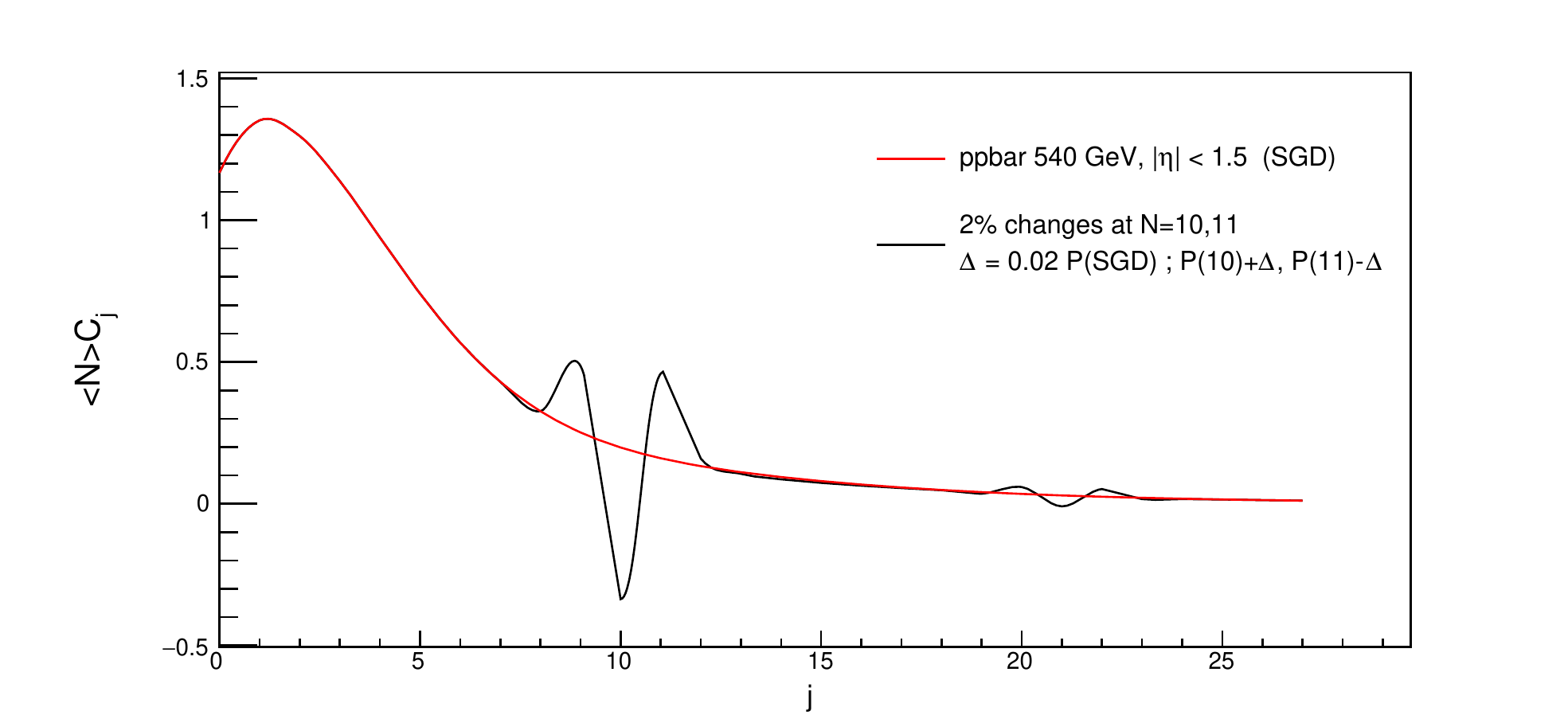} 
\includegraphics[width=4.8 in,height =2.48 in]{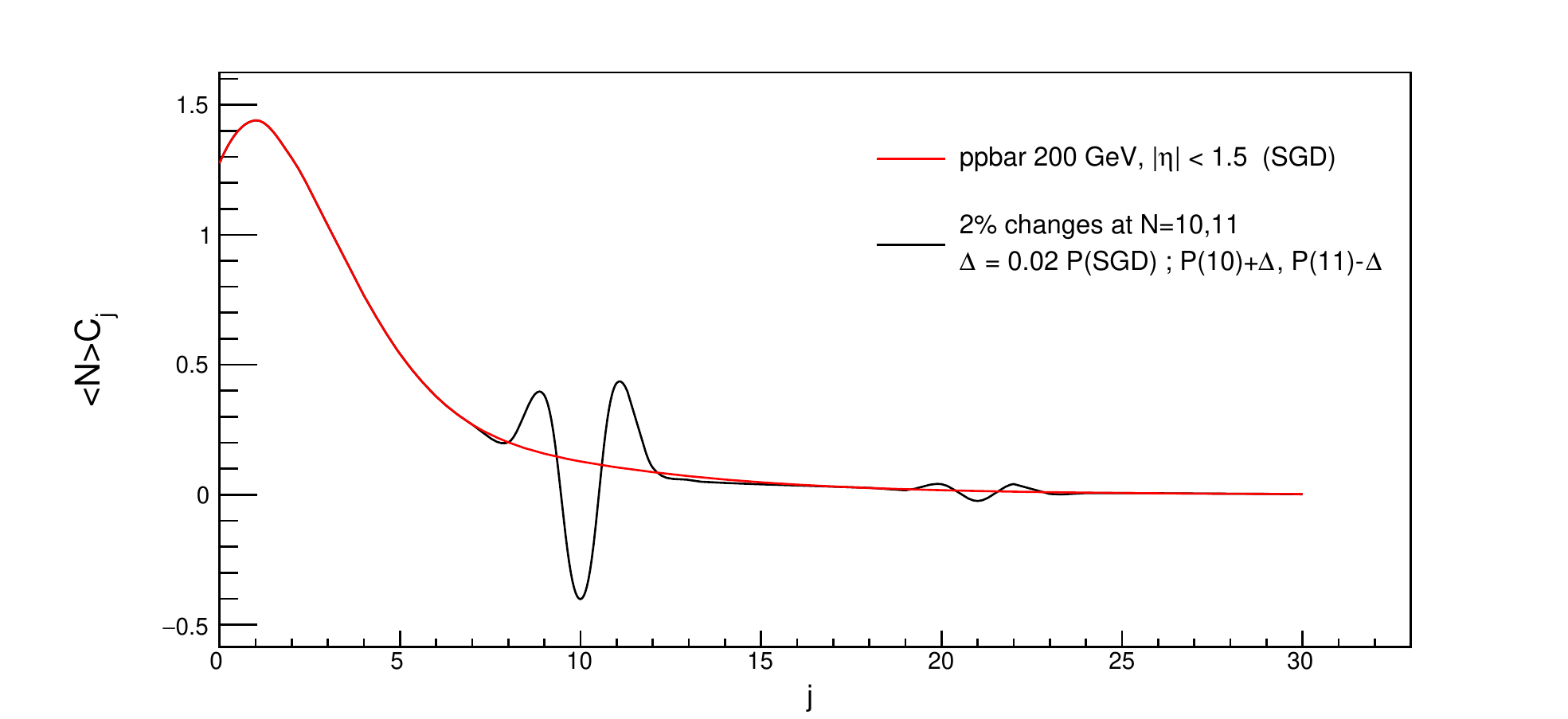} 
\caption{Illustration of oscillatory behaviour of the coefficients $C_j$, as described in the text.}
\end{figure}

The coefficients $C_j$ evaluated from equation~(9) depend on $P(0)$.~In the experimental data from complex detectors, such as CMS at the LHC, the probability $P(0)$ is very large as compared to $P(1)$.~Due to large experimental uncertainties associated with this bin, $P(0)$ is often omitted for the conventional fits to the data.~However $P(0)$ is the only bin which is very sensitive to the acceptance as explained in reference \cite{WILK}.~To show the sensitivity to the value of $P(0)$, we show in figure~10, the coefficients calculated by using the values $P(0)\pm \delta$ for the $pp$ data at $\sqrt{s}$ = 900~GeV for $|\eta|<$ 2.4, where $\delta$ is the error on $P(0)$ measurement.~The coefficients vary with different periods of oscillations, around the values calculated from $P(0)$, as shown in the figure.~Figure~11 shows the oscillatory behaviour when $P(0)$ is not considered, the $C_j$ are calculated starting with $P(1)$.~Coefficients $C_j$ still show the oscillatory behaviour but with much reduced oscillation amplitude, with oscillations dying out quickly.

In equation~(9), the coefficients $C_j$ connect each probability, with every other probability. ~For example $P(N + 1)$ connects to $P(N - j)$, the probabilities of particles produced earlier.~The most important feature of this recurrence relation is that $C_j$ can be directly calculated from the experimentally measured $P(N)$.~In an interesting case study, starting with the SGD, we make changes in successive probabilities by 2$\%$: we put $P(10)$= $P_{SGD}(10) + \Delta$ and $P(11)$ = $P_{SGD}(11) - \Delta$ with $\Delta$ = 0.02$P_{SGD}$ and study the variation of $C_j$ as a function of $j$.~The results are shown in figure~12 for $pp$ collisions at different energies but within the same $|\eta|$ bin.~Similarly, figure~13 shows the plots for $p\overline{p}$ collisions at $\sqrt{s} = $900, 540 and 200~GeV for $|\eta|<$ 3.~The apparently insignificantly small changes in probability, resulted in rather dramatic spikes occurring on the original $P_{SGD}$ and with rapidly falling amplitudes.~This points to the sensitivity of the coefficients $C_{j}$.~Such a change is then provided by the MSGD, whereby spike influences then, the consecutive coefficients $C_{j}$ and brings them to agreement with those obtained from the experimentally measured $P(N)$.~With increasing value of $j$, smaller are the values of $C_j$ and hence weakly influencing the final distribution.~Such behaviour strongly indicates that particles are produced in clusters.

\section{Conclusion}
In this paper we show and reaffirm that the MDs possess a fine structure which can be detected experimentally and analysed in terms of a suitable recurrence relation, such as the one in equation~(9).~The coefficients $C_j$ in the recurrence relation, which are directly connected with the combinants, give a compelling evidence that  phenomenon of oscillatory behaviour of the modified combinants exists in the experimental data on multiplicities.~The coefficients $C_{j}$ have been calculated from the shifted Gompertz distribution and its modified forms; weighted superposition of 2-component shifted Gompertz parametrizations and modified shifted Gompertz distributions including non-linearity to two different orders, equations~(7,8).~The shifted Gompertz distribution, which we introduced in our publication \cite{shGomp}, does not show any oscillatory behaviour.~However its modified forms show the oscillatory behaviour and agree with the data very well.~The oscillations are large at low multiplicities for the $pp$ data and tend to die out at large multiplicities.~In case of $p\overline{p}$ collisions, the oscillations follow a reverse pattern.~The behaviour of oscillations observed in present studies is very similar to what is observed in the case of negative binomial distribution~(NBD), by the authors who pioneered the concept.
\section{Data Availability}
All the data used in the paper can be obtained from the references quoted or from the authors.
\section{Conflicts of Interest}
The author(s) declare(s) that there is no conflict of interest regarding the publication of this article. 
\section*{Acknowledgement}
The author R.~Aggarwal is grateful to the Department of Science and Technology, Government of India for the Inspire-faculty grant. 
\section*{References}
\bibliography{mybibfile}

\begin{enumerate}

\bibitem{KNO}{Z. Koba, H. B. Nielsen, and P. Olesen, "Scaling of multiplicity distributions in high energy hadron collisions," Nuclear Physics B, vol. 40, pp. 317–334, 1972.}

\bibitem{POI}{M. Althoff et al. "Jet Production and Fragmentation in $e^+e^-$ Annihilation at 12-GeV to 43-GeV TASSO Collaboration, Zeitschrift f\"{u}r Physik C: Particles and Fields, vol. 22, pp. 307-340, 1984;  M. Derrick et al. "Study of quark fragmentation in $e^+e^-$ annihilation at 29 GeV: Charged-particle multiplicity and single-particle rapidity distributions," Physical Review D, vol. 34, pp. 3304-3320, 1986.}

\bibitem{NBD} {P. Carruthers and C.C Shih, "The phenomenological  analysis of hadronic multiplicity distributions," International Journal of  Modern Physics A, vol. 2, no. 5, pp. 1447-1497, 1987;  A. Giovannini and R. Ugoccioni, "Negative binomial multiplicity distributions in high energy hadron collisions," Zeitschrift f\"{u}r Physik C:  Particles and Fields, vol. 30, pp. 391-400, 1986 ;"Clan structure analysis and QCD parton showers in multiparticle dynamics: an intriguing dialog between theory and experiment," International Journal of Modern Physics A, vol. 20, no. 17, pp. 3897–3999, 2000, 27th International Symposium on Multiparticle Dynamics (ISMD 97), 8–12 Sep. Frascati, Italy, 1997.} 

\bibitem{LOG} {R. Szwed, G. Wrochna and A. K. Wr\'{o}blewski, "Genesis of the Lognormal multiplicity distribution in the $e^+e^-$ collisions and other stochastic processes," Modern Physics Letters, vol. A, no. 5, pp. 1851-1869, 1990.}

\bibitem{TS1}{C. Tsallis, "Possible generalization of Boltzmann-Gibbs statistics," Journal of Statistical Physics, vol. 52, no. 1-2, pp. 479-487, 1988.}

\bibitem{TS2}{C.E. Ag\"{u}iar and T. Kodama, "Nonextensive statistics and multiplicity distribution in hadronic collisions", Physica A, vol. 320, pp. 371-386, 2003.}


\bibitem{WEI}{S. Dash, B. K. Nandi and P. Sett "Multiplicity distributions in $e^+e^-$ collisions using Weibull distribution", Physical Review, vol. D 94, 074044-49, 2016.}

\bibitem{UA51}{R. E. Ansorge, B. Asman, L. Burow et al., "Charged particle multiplicity distributions at 200 and 900~GeV  cm energy," Zeitschrift f\"{u}r Physik C: Particles and Fields, vol. 43, no. 3, pp. 357–374, 1989.}

\bibitem{UA52} {G. J. Alner, K. Alpg\aa rd, P. Anderer et al., "Multiplicity distributions in different pseudorapidity intervals at a CMS energy of 540 GeV," Physics Letters B, vol. 160, no. 1–3, pp. 193–198, 1985.}

\bibitem{ZBO}{I.J. Zborovsk\'{y},  "A three-component description of multiplicity distributions in pp collisions at the LHC," Journal of Physics G, vol. 40, no. 5, 055005, 2013.}

\bibitem{ZBO2}{I.J. Zborovsk\'{y}, "Three-component multiplicity distribution, oscillation of combinants and properties of clans in pp collisions at the LHC," European Physical Journal C, vol. 78, pp. 816, 2018.}

\bibitem{BEMA}{A.C. Bemmaor. In book by G. Laurent, G. L. Lilien, and B. Pras, Eds.,  "Research Traditions in Marketing," vol. 201, Springer, Netherlands, 1994.}

\bibitem{Jon}{Dragan Juk\'{i} and Darija Markov\'{i},  "Nonlinear least squares estimation of the shifted Gompertz distribution," European Journal of Pure and Applied Mathematics, vol. 10, no. 2, pp. 157–166, 2017.}

\bibitem{Jod}{F. Jim\'{e}nez and P. Jodr\'{a}, "A note on the moments and computer generation of the shifted Gompertz distribution, "  Communications in Statistics, Theory Methods, vol. 38, no. 1, pp. 75–89, 2009.}

\bibitem{Jim}{F.J. Torres, "Estimation of parameters of the shifted Gompertz distribution using least squares, maximum likelihood and moments methods," Journal of Computational and Applied Mathematics, vol. 255, pp. 867–877, 2014.}

\bibitem{shGomp}{Ridhi Chawla and M. Kaur,  "A new distribution for multiplicities in leptonic and hadronic collisions at high energies," Advances in High Energy Physics, vol. 2018, Article ID 5129341, p. 12, 2018.}

\bibitem{AM}{Aayushi Singla and M. Kaur, "Multiplicity moments of shifted Gompertz distribution in $e^+e^-$, $p\overline{p}$ and $pp$ collisions at high energies,"  arXiv:1903.06884v2, 11 Sep. 2019, paper submitted to Advances in High Energy Physics 2019.}

\bibitem{WILK}{G. Wilk and Z. W{\l}odarczyk, "Some intriguing aspects of multiparticle production processes, " International Journal of Modern Physics A, vol. 33, pp. 183008, 2018;  "How to retrieve additional information from the multiplicity distributions," Journal of Physics G: Nuclear and Particle physics, vol. 44, pp. 015002-17, 2017.}

\bibitem{CMS}{V. Khachatryan, A. M. Sirunyan et al., CMS Collaboration, "Charged particle multiplicities in $pp$ interactions at $\sqrt{s}$ = 0.9, 2.36\textbf{•} and 7~TeV," Journal of High Energy Physics, vol. 2011, no. 79, 2011.}

\bibitem{Ryb}{Maciej Rybczy\'{n}ski, Grzegorz Wilk and Zbigniew W{\l}odarczyk,"Intriguing properties of multiplicity distributions", Physical Review D, vol. 99, pp. 094045, 1-10, 2019.}

\bibitem{Abram}{V. A. Abramovsky and N. V. Radchenko,"Multiplicity Distribution in Proton–Proton 
and Proton–Antiproton Collisions at High Energies,"Physics of Particles and Nuclei Letters, Vol. 6, No. 6, pp. 433–439, 2009.} 

\bibitem{Ang}{H.W. Ang, A.H. Chan, M. Ghaffar, M. Rybczy\'{n}ski, G. Wilk and Z. W{\l}odarczyk,"A look at multiparticle production via modified combinants ,"arXiv:1908.11062v1 [hep-ph] 29 Aug 2019.} 
 

\bibitem{ALICE}{J. Adam et al., ALICE Collaboration, "Charged-particle multiplicity distributions over a wide pseudorapidity range in proton-proton collisions at $\sqrt{s}$=0.9, 7 and 8~TeV, " European Physical Journal C, vol. 77, pp. 852-885, 2017.}
\end{enumerate}

\end{document}